\documentclass[
  aps,prd,twocolumn,superscriptaddress,nofootinbib,longbibliography,floatfix
]{revtex4-2}

\usepackage{graphicx}
\usepackage{amsmath,amssymb}
\usepackage{bm}
\usepackage{booktabs}
\usepackage{xcolor}
\usepackage{hyperref}
\hypersetup{colorlinks=true,linkcolor=blue,citecolor=blue,urlcolor=blue}

\newcommand{\Msun}{M_\odot}
\newcommand{\castor}{\textsc{Castor}}
\newcommand{\gww}{GW-Whisper}
\newcommand{\Dsens}{D_{\mathrm{sens}}}

\begin{document}

\title{A Fast and Scalable Transformer Pipeline for Binary Black Hole Detection}

\author{Chayan Chatterjee}
\affiliation{Department of Physics and Astronomy, Vanderbilt University\\ 2201 West End Avenue, Nashville, Tennessee - 37235,}
\affiliation{Data Science Institute, Vanderbilt University\\ 1400 18th Avenue South Building, Suite 2000, Nashville, Tennessee - 37212,}
\email{chayan.chatterjee@vanderbilt.edu}

\author{Abigail Petulante}
\affiliation{Data Science Institute, Vanderbilt University\\ 1400 18th Avenue South Building, Suite 2000, Nashville, Tennessee - 37212,}
\email{abigail.petulante@vanderbilt.edu}

\author{Haowei Fu}
\affiliation{Data Science Institute, Vanderbilt University\\ 1400 18th Avenue South Building, Suite 2000, Nashville, Tennessee - 37212,}
\email{haowei.fu@vanderbilt.edu}

\author{Yang Hu}
\affiliation{Data Science Institute, Vanderbilt University\\ 1400 18th Avenue South Building, Suite 2000, Nashville, Tennessee - 37212,}
\email{yang.hu.1@vanderbilt.edu}

\author{Roy Lau}
\affiliation{Data Science Institute, Vanderbilt University\\ 1400 18th Avenue South Building, Suite 2000, Nashville, Tennessee - 37212,}
\email{roy.lau@vanderbilt.edu}

\author{Karan Jani}
\affiliation{Department of Physics and Astronomy, Vanderbilt University\\ 2201 West End Avenue, Nashville, Tennessee - 37235,}
\email{karan.jani@vanderbilt.edu}

\date{\today}

\begin{abstract}

With the projected increase in the detection rate of compact-binary coalescences in the coming decade, there is critical need to develop fast, robust, and scalable alternatives to matched filtering for gravitational-wave searches. Transformer models have revolutionized natural language and audio processing but their application to gravitational-wave astronomy is still largely unexplored. In this work, we introduce \castor, a transformer-based coincident search pipeline for detecting binary black hole gravitational-wave signals from Advanced LIGO detectors. One of the major features of our model is that it allows the false-alarm rate to be estimated via time slides cheaply without requiring repeated evaluations of the neural network. We evaluate \castor\ on datasets from the Machine-Learning Gravitational-Wave Search Challenge (MLGWSC-1) and on approximately five months of real O3b observing strain. When tested on benchmark datasets, \castor\ ranks among the most sensitive machine-learning pipelines and successfully recovers the majority of confident events from the GWTC-3 catalog that lie within its training range. We also benchmark \castor\ against another transformer architecture, GW-Whisper, a domain-adaptation of OpenAI's audio foundation model. We find that \castor\ substantially outperforms the repurposed audio model in sensitivity and also reduces the computational cost of background estimation by a factor of 20. Our results demonstrate a highly practical, scalable approach for deep-learning gravitational wave searches and empirical background estimation for future observing runs.

%The projected increase in the detection rate of compact-binary coalescences motivates fast, robust, and scalable alternatives to matched filtering for gravitational-wave searches. While transformer models have revolutionized natural language and audio processing, their application to gravitational-wave astronomy remains largely unexplored. We introduce \castor, a fast, transformer-based pipeline for detecting binary black holes in Advanced LIGO data. By processing data from each detector independently before combining the results, \castor\ sidesteps a major computational bottleneck in machine-learning searches: it allows the false-alarm rate to be estimated via time slides without requiring repeated evaluations of the neural network. We evaluate \castor\ on datasets from the Machine-Learning Gravitational-Wave Search Challenge (MLGWSC-1) and on approximately five months of real O3b observing strain. \castor\ ranks among the most sensitive machine-learning pipelines on benchmark data and successfully recovers the majority of confident GWTC-3 catalog events within its training range. Furthermore, we benchmark \castor\ against GW-Whisper, an adaptation of OpenAI's audio foundation model. We find that our purpose-built, time-domain network substantially outperforms the repurposed audio model in sensitivity, while our coincident architecture reduces the computational cost of background estimation by a factor of 20. Our results demonstrate a highly practical, scalable approach for deep-learning gravitational wave searches and empirical background estimation.

\end{abstract}

\maketitle

% ======================================================================
\section{Introduction}
\label{sec:intro}
% ======================================================================

The Advanced LIGO--Virgo--KAGRA (LVK) network~\cite{LIGO, Virgo, KAGRA_1} has transformed gravitational-wave (GW) astronomy. Since the first direct detection in 2015 from a binary black hole (BBH) merger~\cite{Abbott2016GW150914}, the discoveries of compact-binary coalescences (CBCs) has grown rapidly. The fourth Gravitational Wave Transient Catalog (GWTC-4.0) added 128 candidates from the first part of the fourth observing run (O4a) with probability of astrophysical origin $p_{\rm astro}\geq 0.5$, bringing the cumulative detections to 218 candidates~\cite{GWTC4,GWTC4Methods}. Further reanalysis updated the number of candidates to 229, and GWTC-5.0 added 161 candidates from the second half of the fourth observing run, O4b, yielding a total of 390 events with $p_{\rm astro}\geq 0.5$ ~\cite{GWTC5}. Planned sensitivity upgrades and next-generation observatories are expected to increase the detection rate to several events per day~\cite{Reitze2019,Punturo2010,Maggiore2020}. This will place enormous demands on both the low-latency and offline search pipeline infrastructures that identify, characterize and candidate events.

%The first direct detection of gravitational waves (GWs) from a binary black-hole (BBH) merger by the Advanced LIGO detectors~\cite{Aasi2015,Abbott2016GW150914} inaugurated GW astronomy. The Advanced LIGO--Virgo--KAGRA network~\cite{Acernese2015,Akutsu2021} has since transformed compact-binary coalescences (CBCs) from individual discoveries into a rapidly growing population.  GWTC-4.0 added 128 candidates from the first part of the fourth observing run (O4a) with probability of astrophysical origin $p_{\rm astro}\geq 0.5$, bringing the cumulative catalog to 218 candidates~\cite{GWTC4,GWTC4Methods}.  A subsequent reanalysis updated the number of candidates through O4a to 229, and GWTC-5.0 added 161 candidates from O4b, yielding 390 catalogued transients with $p_{\rm astro}\geq 0.5$ through the end of O4b~\cite{GWTC5}.  The increasing event rate is accompanied by a growing dynamic range: GWTC-4.0 reported the first BBH signals with network signal-to-noise ratio (SNR) above 30 and included GW231123, the most massive BBH merger in the catalog at that time, while GWTC-5.0 contains five additional BBH signals with network SNR above 30 and a maximum network SNR of 76.9~\cite{GWTC4,GWTC5}.  Planned sensitivity upgrades and next-generation observatories are expected to raise the detection rate to several events per day~\cite{Reitze2019,Punturo2010,Maggiore2020}, placing increasing demands on the low-latency and offline pipelines that identify, rank, validate, and characterize candidate events.

The optimal method for detecting modelled CBC signals under the assumption of known signal morphology and stationary, Gaussian noise is matched filtering~\cite{Allen2012,Usman2016,Nitz2017,Sachdev2019,Aubin2021,SPIIR}, in which detector data are cross-correlated against banks of template waveforms~\cite{Owen1999,Roy2019}. In actual LVK searches, template-bank filtering is combined with signal-consistency tests, empirical background estimation, data-quality verification, and event validation to suppress non-Gaussian and non-stationary noise transients or ``glitches'' present in real data~\cite{Allen2005,Nitz2018,Davis2021,GWTC4Methods, LIGO_DetChar_O4_b_c, LIGO_DQR}. The computational cost of this search method grows as they cover broader source parameter spaces and extend to lower frequencies. Recent work has therefore also modernized conventional searches: a PyTorch implementation of the GstLAL search pipeline retained comparable search performance on public data while providing substantial GPU acceleration~\cite{Huang2025Scalable}. A complementary, weakly modelled search is performed using the coherent WaveBurst (cWB) pipeline~\cite{Klimenko2016,Drago2021} that searches for coherent excess power and is particularly valuable for signals that are poorly represented by available template banks.

%The standard detection method for modelled CBC signals is matched filtering~\cite{Allen2012,Usman2016,Nitz2017,Sachdev2019,Aubin2021}, in which detector data are cross-correlated against banks of template waveforms~\cite{Owen1999,Roy2019}. Matched filtering is optimal when the signal is known and the detector noise is stationary and Gaussian.  In practice, production searches combine template-bank filtering with signal-consistency tests, coincidence or coherence across detectors, empirical background estimation, data-quality information, and event validation to suppress the non-Gaussian, non-stationary noise transients (``glitches'') present in real data~\cite{Allen2005,Nitz2018,Davis2021,GWTC4Methods}.  Their computational cost grows as searches cover broader source parameter spaces and extend to lower frequencies, where inspiral signals remain in band for longer.  Recent work has therefore also modernized conventional searches: a PyTorch implementation of the GstLAL filtering engine retained comparable search performance on public data while providing substantial GPU acceleration~\cite{Huang2025Scalable}. The coherent WaveBurst (cWB) pipeline~\cite{Klimenko2016,Drago2021} provides a complementary, weakly modelled search for coherent excess power and is particularly valuable for signals that are poorly represented by available template banks.

Machine learning, and in particular, deep learning has emerged as a faster alternative to traditional GW data-analysis methods, and is under active development for many applications, including rapid detection~\cite{George2018,Gabbard2018, Gebhard2019,Krastev2020,Schafer2020,Wang2020, Beveridge2025, Beveridge_BBH, Yamamoto2025,Malz2026}, parameter estimation~\cite{Chua2020,Green2021,Dax2021,GW-SkyLocator, GW-SkyLocator_pre-merger}, and glitch classification~\cite{Zevin2017,George2018Glitch,Bahaadini2018,AST} (see Ref.~\cite{Cuoco2025Review} for a recent review). The advantage here is that neural networks can amortize the cost of searches or expensive Bayesian analyses into a one-time training stage, and then use the trained model for inference on new data at low computational cost and latency.  A common benchmark to test the performance of deep learning models for end-to-end BBH searches in simulated Gaussian noise and real detector noise is the first Machine-Learning Gravitational-Wave Search Challenge (MLGWSC-1)~\cite{Schafer2023}, established to test direct comparisons of machine learning searches (MFCNN~\cite{Wang2020}, CNN-Coinc, the TPI-FSU Jena convolutional search~\cite{Zelenka2024}, and the AResGW/Virgo-AUTh residual network~\cite{Nousi2023}), against representative PyCBC \cite{Usman2016, Nitz2017} and cWB \cite{Klimenko2016,Drago2021} pipelines. Such benchmarks are essential because classification accuracy or receiver operating characteristic curves alone do not measure search performance at the very low false-alarm rates required for astrophysical discovery.

Recent studies have also moved data-driven or learned GW detection beyond short, independently classified data segments toward continuous, multi-detector searches with empirically estimated backgrounds. A fully machine-learning-based low-latency pipeline has demonstrated state-of-the-art sensitivity to higher-mass stellar BBHs with inference latency below that of conventional matched-filter pipelines~\cite{Marx2025Aframe, Aframe}.  Deep-learning searches for binary neutron-star mergers have also achieved sensitivity comparable to established offline searches in real data \cite{Aframe_BNS, McLeod2025BNS}. AResGW has been applied to archival O1--O3 data and reported new candidate events identified by a machine-learning search~\cite{Koloniari2025AresGW}. Machine learning is additionally being explored as a way to combine the outputs of heterogeneous search pipelines while retaining calibrated uncertainty estimates~\cite{Ashton2026Combination}.

%Deep learning has emerged as a fast alternative for several GW data-analysis tasks, including rapid detection~\cite{George2018,Gabbard2018, Gebhard2019,Krastev2020,Schafer2020,Wang2020}, parameter estimation~\cite{Chua2020,Green2021,Dax2021,Chatterjee2021}, and glitch classification~\cite{Zevin2017,George2018Glitch,Bahaadini2018}; see Ref.~\cite{Cuoco2025Review} for a recent review.  Neural networks can amortize much of the cost of waveform recognition into a one-time training stage and then evaluate new data with low latency.  The Machine-Learning Gravitational-Wave Search Challenge (MLGWSC-1)~\cite{Schafer2023} established a common benchmark for end-to-end searches in simulated Gaussian coloured noise and real detector noise.  It enabled direct comparisons of learned pipelines, including MFCNN~\cite{Wang2020}, CNN-Coinc, the TPI-FSU Jena convolutional search~\cite{Zelenka2024}, and the AResGW/Virgo-AUTh residual network~\cite{Nousi2023}, with representative PyCBC \cite{Usman2016, Nitz2017} and cWB \cite{Klimenko2016,Drago2021} configurations. Such benchmarks are essential because classification accuracy or receiver operating characteristic curves alone do not measure search performance at the very low false-alarm rates required for astrophysical discovery.

These advances also make it necessary to develop credible learned searches. The performance of machine learning-based approaches can appear strong without robust generalization to new observing conditions~\cite{sage,Cuoco2025Review}. This is because sensitivity estimates can depend strongly on the training prior, waveform family, noise realization, preprocessing choices, etc. A practical pipeline must therefore operate on continuous multi-detector data and demonstrate robustness to evolving noise spectra and out-of-distribution signals. Candidate events must be assigned significance using sufficiently long background livetime to resolve astrophysically relevant false-alarm rates. Meeting these requirements while preserving the speed of learned inference is a central challenge for machine-learning-based GW detection.

This work is motivated by a central trade-off in multi-detector searches. Traditionally, information from multiple detectors is combined using either a coherent or a coincident approach. Coherent searches analyze the detector data jointly, enabling a model to identify cross-detector consistency and potentially achieve greater sensitivity. The drawback is the substantial computational cost of estimating the background. Establishing a reliable false-alarm rate (FAR) at the $1,\mathrm{month}^{-1}$ level for online searches—or $1,\mathrm{year}^{-1}$ for offline searches—requires months to years of effective background livetime. For a two-detector network, this background is typically constructed using time slides, in which the data from one detector are repeatedly shifted relative to the other by more than the inter-site light-travel time. This removes genuine astrophysical coincidences, so any remaining coincident events can be treated as false alarms~\cite{Was2010,Usman2016}. When a neural network processes the detector strains jointly, however, the model must be rerun for every time shift, making large-scale background estimation prohibitively expensive.

Coincident searches offer a computationally efficient alternative. Each detector is analyzed independently, and the resulting single-detector outputs are subsequently combined. The network therefore needs to be evaluated only once for each detector, while time slides can be generated inexpensively by recombining cached outputs. This scalable strategy is widely used in production matched-filter searches~\cite{Usman2016,Sachdev2019}, but it may not fully exploit the sensitivity gains available from coherent cross-detector information. The challenge, therefore, is to retain the representational power and cross-detector awareness of a coherent search \cite{SPIIR, Klimenko2016, Drago2021} while preserving the computational efficiency of a separable, post-hoc ranking statistic. Transformer networks are particularly well suited to this goal because their attention mechanisms provide a powerful framework for learning structured relationships within sequential data.

In this paper, we introduce \castor\ (Coincident Analysis Siamese TransfORmer), a fast and sensitive transformer-based search for BBH signals designed to overcome this computational limitation. \castor\ is a compact time-domain model, containing approximately $2.9$ million parameters, that is trained from scratch and applied separately to each detector. The detector streams are processed using a shared-weight Siamese architecture, and the resulting single-detector detection statistics are combined to construct a coincident ranking statistic. Because the neural network is evaluated only once for each detector, time-slide backgrounds can be generated directly from the cached outputs. Each slide therefore requires only the inexpensive recomputation of the closed-form coincident statistic, without any additional neural-network inference. Our aim is to demonstrate a fast, accurate time-domain transformer BBH search and validate a statistically well-calibrated time-slide background that requires no additional network evaluations. We evaluate \castor\ against a separate transformer-based model, \gww~\cite{Chatterjee2024GWWhisper}, a domain-adapted version of OpenAI's Whisper audio foundation model~\cite{Radford2023}, pre-trained on 680,000 hours of human speech data. Audio foundation models provide a promising starting point because both human speech and GW signals contain structured time--frequency features and lie in the same frequency range. Unlike \castor\, \gww\ jointly processes $Q$-transforms of single-detector data and is fine-tuned using low-rank adapters, which trains a small subset of model parameters to GWs while retaining the audio representations it was pre-trained on \cite{Liu2024DoRA}. This comparison allows us to directly assess the trade-offs between a specialized, time-domain transformer trained from scratch and a repurposed, transfer-learned audio foundation model operating on time--frequency representations.

The remainder of the paper is organised as follows. Section~\ref{sec:theory} reviews the detection statistic, and other figures of merit.  Section~\ref{sec:methods} describes the two networks, their training, ranking statistics, and the time-slide procedure.  Section~\ref{sec:data} describes the datasets used for evaluating the networks. Section~\ref{sec:results} presents results on the MLGWSC-1 datasets and search over real O3b data. Finally, Section~\ref{sec:discussion} discusses the implications of the results and plans for future work.

% ======================================================================
\section{Detection statistic and figures of merit}
\label{sec:theory}
% ======================================================================

\subsection{Matched filtering and the optimal statistic}

Suppose the data contain a known signal $h(t)$ together with additive noise $n(t)$, such that $d(t)=h(t)+n(t)$. For stationary Gaussian noise, the optimal Neyman--Pearson detection statistic is the matched-filter signal-to-noise ratio (SNR)~\cite{Allen2012}. We define the noise-weighted inner product as
\begin{equation}
\langle a|b\rangle = 4\,\mathrm{Re}\int_{f_\mathrm{low}}^{f_\mathrm{high}}
\frac{\tilde a(f)\,\tilde b^{*}(f)}{S_n(f)} \,df ,
\label{eq:inner}
\end{equation}
where $S_n(f)$ is the one-sided power spectral density (PSD) of the noise. The matched-filter SNR for a template $h$ is then given by $\rho=\langle d|h\rangle/\sqrt{\langle h|h\rangle}$.

Learned searches take a different approach: instead of explicitly correlating the data with a template, they learn a function $\mathcal{R}(d)$ from labelled examples of signals and noise. The network output can be used to rank candidate events. Its statistical significance must be determined empirically by comparing it with triggers drawn from the noise background. For a two-detector search, this also requires combining the individual detector outputs. The coincident statistic used in this work is described in Sec.~\ref{sec:usr}.

\subsection{False-alarm rate}

For a ranking-statistic threshold $R$, the false-alarm rate is the rate at which noise produces events with $\mathcal{R}(d)\ge R$:
\begin{equation}
  \mathrm{FAR}(R) = \frac{N_\mathrm{bg}(\mathcal{R}(d) \ge R)}{T_\mathrm{bg}},
  \label{eq:far}
\end{equation}
where $N_\mathrm{bg}\left(\mathcal{R}(d)\ge R\right)$ is the number of accidental background events at or above the threshold and $T_\mathrm{bg}$ is the total background live time~\cite{Usman2016,Zelenka2024}. In practice, this background is estimated using time slides described earlier. The total background live time is the sum of the analyzed coincident duration across all slides and can therefore be much longer than the unshifted, or zero-lag, observation time.

If a candidate is louder than every event in the measured background, then $N_\mathrm{bg}=0$ and its empirical FAR would formally be zero. Since a finite background cannot establish a truly zero event rate, we adopt the standard plus-one prescription, $\mathrm{FAR}(R)=[1+N_\mathrm{bg}(\ge R)]/T_\mathrm{bg}$. The loudest candidates are therefore assigned the smallest measurable FAR of $1/T_\mathrm{bg}$.

\subsection{Sensitive distance and sensitive volume}

Following the MLGWSC-1 convention~\cite{Schafer2023}, the sensitivity of a search at a fixed FAR $F$ is quantified by the sensitive volume $V(F)$, i.e.\ the space-time volume within which sources are detected, estimated from an injection campaign by
\begin{equation}
  V(F) \;\approx\; V(d_\mathrm{max}) \cdot
  \frac{1}{N_\mathrm{inj}}
  \sum_{i=1}^{N_\mathrm{found}(F)}
  \left(\frac{\mathcal{M}_{c,i}}{\mathcal{M}_{c,\mathrm{max}}}\right)^{5/2},
  \label{eq:volume}
\end{equation}

where $\mathcal{M}_{c,i}$ is the (detector-frame) chirp mass of the $i$-th found injection, $\mathcal{M}_{c,\mathrm{max}}$ is the upper limit of the injected chirp-mass distribution, $N_\mathrm{inj}$ is the total number of injections, $V(d_\mathrm{max})$ is the volume of the sphere out to which injections are placed, and $N_\mathrm{found}(F)$ is the number of injections recovered with FAR $\le F$. The $(\mathcal{M}_c/\mathcal{M}_{c,\mathrm{max}})^{5/2}$ weighting accounts for the chirp-mass dependence of the intrinsic detectable volume. However, as discussed in~\cite{sage}, this scaling inherently biases the aggregate sensitive volume and distance metrics toward a search's sensitivity to high-mass events, meaning the overall metric is dominated by the recovery of high-mass systems. The sensitive distance is the radius of the equivalent sphere,

\begin{equation}
  \Dsens(F) = \left(\frac{3\,V(F)}{4\pi}\right)^{1/3}.
  \label{eq:dsens}
\end{equation}

An injection is counted as found at FAR $F$ if the search produces a clustered trigger within a fixed time window of the injection time whose ranking statistic corresponds to a FAR $\le F$~\cite{Schafer2023}. The details of the clustering and evaluation method are described in Sec.~\ref{sec:castor}. All results use Eqs.~\eqref{eq:volume}--\eqref{eq:dsens} with the MLGWSC-1 evaluation code~\cite{Schafer2023}.

%where $\mathcal{M}_{c,i}$ is the (detector-frame) chirp mass of the $i$-th found injection, $\mathcal{M}_{c,\mathrm{max}}$ is the upper limit of the injected chirp-mass distribution, $N_\mathrm{inj}$ is the total number of injections, $V(d_\mathrm{max})$ is the volume of the sphere out to which injections are placed, and $N_\mathrm{found}(F)$ is the number of injections recovered with FAR $\le F$. The $(\mathcal{M}_c/\mathcal{M}_{c,\mathrm{max}})
%^{5/2}$ weighting accounts for the chirp-mass dependence of the intrinsic detectable volume. The sensitive distance is the radius of the equivalent sphere,

%\begin{equation}
%  \Dsens(F) = \left(\frac{3\,V(F)}{4\pi}\right)^{1/3}.
%  \label{eq:dsens}
%\end{equation}

%An injection is counted as found at FAR $F$ if the search produces a clustered trigger within a fixed time window of the injection time whose ranking statistic corresponds to a FAR $\le F$~\cite{Schafer2023}. The details of the clustering and evaluation method are described in Sec.~\ref{sec:castor}. All results use Eqs.~\eqref{eq:volume}--\eqref{eq:dsens} with the MLGWSC-1 evaluation code~\cite{Schafer2023}.

% ======================================================================
\section{Methods}
\label{sec:methods}
% ======================================================================

\subsection{Transformers and the attention mechanism}
\label{sec:transformer-overview}

Transformers~\cite{Vaswani2017} treat an ordered input as a sequence of finite-dimensional vectors known as tokens. In language models these tokens usually correspond to words or portions of a sentence. For continuous signals they instead encode localized regions or compressed features of the raw data. Each token is mapped to an embedding of width $d_\mathrm{model}$. Since the transformer architecture has no built-in notion of order, we inject spatial or temporal context by adding a positional embedding. The token embedding then encodes what features are present and the positional embedding encodes where they occur.

Self-attention is the core operation of the encoder. Given an embedded sequence $X\in\mathbb{R}^{N\times d_\mathrm{model}}$, the network learns linear projections to the query, key, and value matrices, $Q=XW_Q$, $K=XW_K$, and $V=XW_V$, and computes scaled dot-product attention as
\begin{equation}
\operatorname{Attention}(Q,K,V)
= \operatorname{softmax}\!\left(
\frac{QK^{\mathsf T}}{\sqrt{d_k}}
\right)V,
\label{eq:self-attention}
\end{equation}
with $d_k$, the dimensionality of the queries and keys. The product $QK^{\mathsf T}$ measures the pairwise affinity between every pair of tokens, the softmax turns these affinities into attention weights, and each token's representation is updated as a weighted sum over the whole sequence.

To capture patterns on different scales, such as short transients as well as long-range correlations, this mechanism is split across several attention heads running in parallel. A standard encoder block combines this multi-head attention with a position-wise feed-forward network. Stacking such blocks lets the model progressively enrich each token's local features with broader context.

Before the sequence is passed to a downstream classification or regression task, it usually has to be collapsed to a fixed-size vector. The simplest choice is mean pooling, which averages the contextualized representations,
\begin{equation}
z_{\mathrm{mean}}
= \frac{1}{N}\sum_{i=1}^{N} h_i,
\label{eq:mean_pool}
\end{equation}
where $h_i\in\mathbb{R}^{d_\mathrm{model}}$ is the $i$-th token representation. The attention layers already weight different regions differently, but the averaging step itself treats every token equally, so a sharply localized signal can be diluted by background noise.

Attention pooling avoids this by learning how much each token should contribute. Given a scoring function $a(\cdot)$, the normalized weights and the pooled vector are
\begin{equation}
\alpha_i
= \frac{\exp\!\left(a(h_i)\right)}
{\sum_{j=1}^{N}\exp\!\left(a(h_j)\right)},
\qquad
z_{\mathrm{attn}}=\sum_{i=1}^{N}\alpha_i h_i,
\label{eq:attention_pool}
\end{equation}
so that the summary vector is a learned weighted combination of the tokens rather than a flat average. Decoupling the weights from a strict average lets the model emphasize salient localized features and suppress noisy or irrelevant segments. An alternative is to prepend a learnable summary token to the sequence and read out its final state. The pooled vector is then fed to a task-specific head that produces class logits, regression targets, or continuous embeddings. In this work, we retain the full sequence and pass it through a separate frame-level head, so the network outputs both a global signal vs. noise classification score and localized per-token prediction at the same time. The details are provided in the next section.

%The two architectures studied here adapt this framework to GW data in different ways. \castor\ tokenizes the raw time-domain strain with strided one-dimensional convolutions and feeds the tokens into a randomly initialized Transformer encoder tuned for GW signals. \gww, by contrast, first projects the strain into a time--frequency representation and processes it with a pretrained Whisper audio encoder. Both rely on Transformer blocks for contextual modeling, but they differ in their tokenization, input domain, and use of pretraining.

\begin{figure*}[t]
\centering
\includegraphics[width=0.92\textwidth]{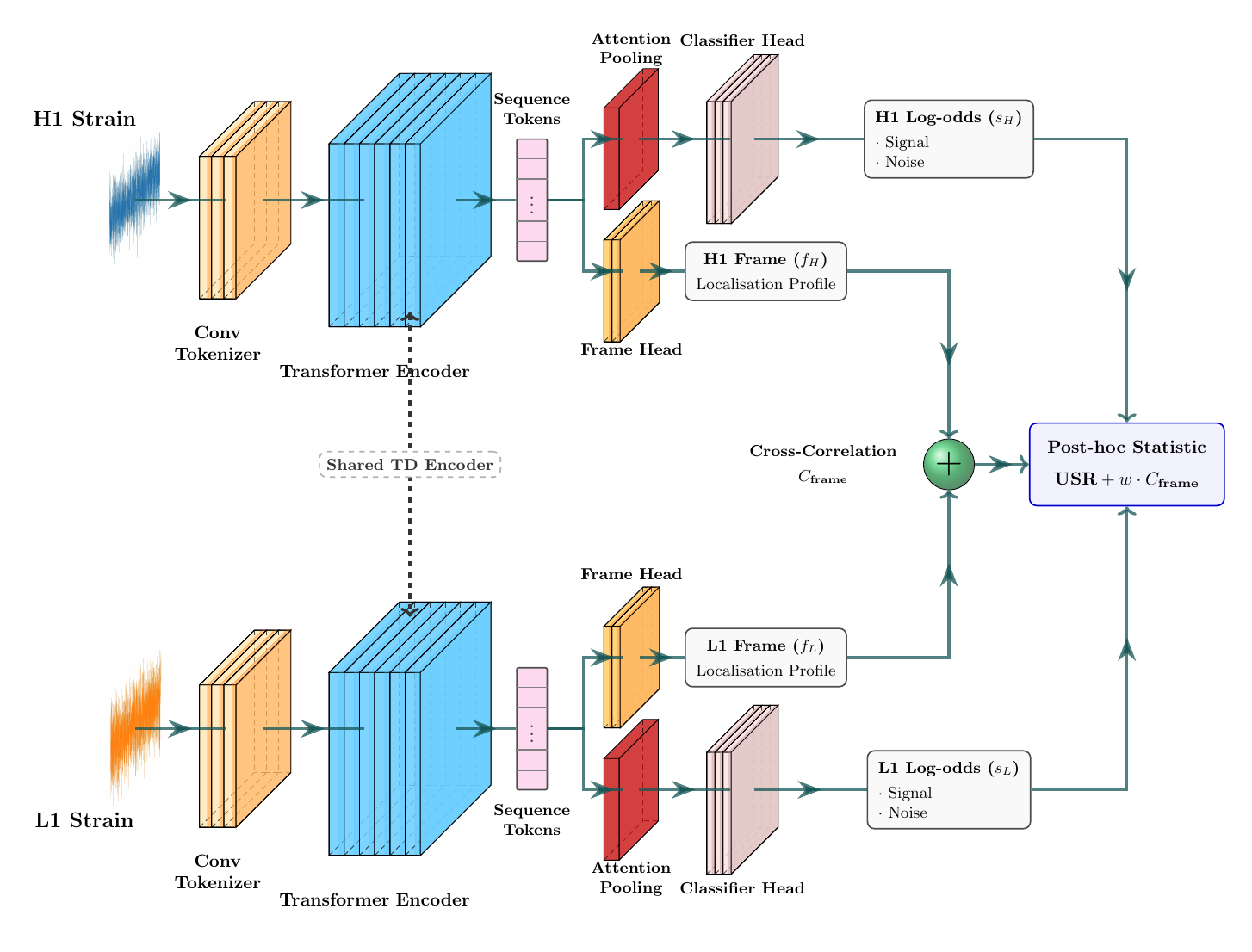}
\caption{Architecture of \castor. The whitened, band-passed strain from H1 (top) and L1 (bottom) is tokenized with strided convolutions, and the two streams pass through a single shared Transformer encoder. Within each branch, attention pooling and a classification head produce a signal-versus-noise log-odds statistic ($s_H$ or $s_L$), while a frame head extracts per-token merger-localization profiles ($f_H$ and $f_L$). The single-interferometer statistics are combined into the unbounded softmax-ratio (USR) statistic, and cross-correlating the frame profiles gives the temporal-coherence term $\mathcal{C}_{\mathrm{frame}}$. The final ranking statistic is $\mathrm{USR}+w_{\mathrm{c}}\mathcal{C}_{\mathrm{frame}}$. Caching the single-detector outputs lets us run rapid time slides without repeated network inference.}
\label{fig:arch}
\end{figure*}

\subsection{\castor: a per-detector time-domain transformer}
\label{sec:castor}

\castor\ operates directly on whitened, calibrated time-domain strain data. For each interferometer, the input consists of a $1\,\mathrm{s}$ duration segment sampled at $2048\,\mathrm{Hz}$ ($2048$ samples). The front end of the network utilizes a stack of three strided 1D convolutional layers to compress the continuous strain into a sequence of learned physical feature representations. The convolutional filters progressively increase the feature dimensionality ($64 \to 128 \to 192$) while reducing the temporal resolution with successive strides of $4$, $4$, and $2$. This yields a temporal downsampling factor of $32$, converting each $1\,\mathrm{s}$ strain segment into $N=64$ contextual tokens of dimension $d_\mathrm{model}=192$, with each token effectively spanning an equivalent duration of $31.25\,\mathrm{ms}$.

The token sequence is augmented with learned positional embeddings and processed by a standard 6-layer Transformer encoder with 6 attention heads per layer, an internal feed-forward dimension of $4d_\mathrm{model}=768$, and layer normalization. In our Siamese configuration, identical network weights are shared across interferometers, evaluating each detector strain stream independently. 

To aggregate the sequence of contextualized representations $\{h_i\}_{i=1}^N$ into a global segment summary vector $z_{\mathrm{attn}}\in\mathbb{R}^{d_\mathrm{model}}$ for classification, we implement a single-query scaled dot-product attention pooling layer. This layer uses a single learnable reference query vector, $q \in \mathbb{R}^{d_\mathrm{model}}$, initialized from $\mathcal{N}(0, 1/d_\mathrm{model})$. The alignment score $e_i$ of each temporal token $h_i$ against this global query is computed via a scaled dot product:

\begin{equation}
e_i = \frac{h_i \cdot q}{\sqrt{d_\mathrm{model}}},
\label{eq:castor_attn_scores}
\end{equation}

yielding the normalized softmax weights $\alpha_i = \exp(e_i)/\sum_{j=1}^{N}\exp(e_j)$. The pooled representation $z_{\mathrm{attn}} = \sum_{i=1}^N \alpha_i h_i$ is then mapped by a two-layer perceptron (MLP) to per-detector two-class logits (signal versus noise). This parametric query dynamically focuses weight on transient signal power (such as the merger chirp) while assigning minimal weight to stationary background noise tokens.

%In parallel with global classification, \castor\ preserves temporal resolution by passing the unpooled token sequence $\{h_i\}_{i=1}^N$ through a lightweight per-token ``frame'' head. This head acts as a local signal-posterior predictor and is trained jointly against a 1D Gaussian target centered at the estimated merger token. As shown in Fig.~\ref{fig:frame}, the predicted per-detector frame profiles closely track the injected merger times in both instruments. Cross-correlating these single-detector temporal profiles provides an explicit cross-detector time-coherence statistic (detailed in Sec.~\ref{sec:usr}), enforcing physical astrophysical coincidence across the detector network. The total model contains $\sim 2.9\times 10^6$ trainable parameters. \\

\begin{figure*}[t]
  \centering
  \includegraphics[width=\textwidth]{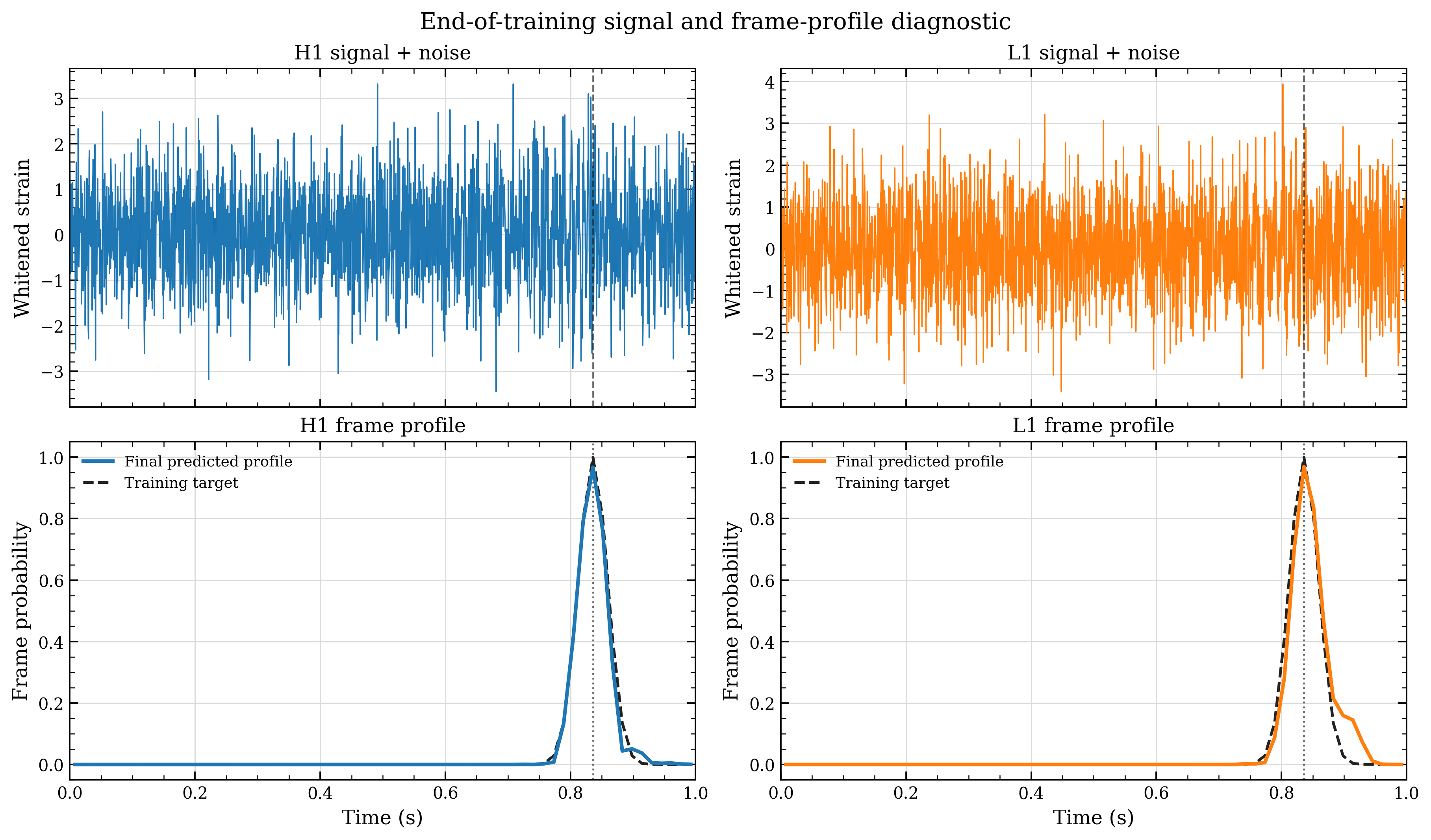}
  \caption{Diagnostic output of the \castor\ frame head on an injected BBH event. \textbf{Top:} Whitened $1\,\mathrm{s}$ strain time series in H1 (left) and L1 (right) with the injected merger epoch indicated (dashed vertical line). \textbf{Bottom:} Corresponding single-detector predicted frame profiles $f_d$ (solid curves) compared against the target Gaussian merger localization (dashed curves). The target is constructed by centering a Gaussian profile on the token containing the peak amplitude of the noise-free signal. Both predicted profiles peak synchronously at the true merger token, providing the temporal coincidence required to maximize the coherence statistic in Eq.~\eqref{eq:framecoherence}.}
  \label{fig:frame}
\end{figure*}

In parallel with global classification, \castor\ preserves temporal resolution by passing the unpooled token sequence $\{h_i\}_{i=1}^N$ through a lightweight per-token ``frame'' head. This head acts as a local signal-posterior predictor and is trained jointly against a 1D Gaussian target profile. This target is constructed during training by locating the temporal token that contains the peak amplitude of the injected, noise-free whitened waveform, and centering a Gaussian distribution over the token sequence at that index. Providing a smooth Gaussian target, rather than a single-token impulse, penalizes the network less for near-misses and encourages it to smoothly localize the merger time. As shown in Fig.~\ref{fig:frame}, the predicted per-detector frame profiles closely track the injected merger times in both instruments. Cross-correlating these single-detector temporal profiles provides an explicit cross-detector time-coherence statistic (detailed in Sec.~\ref{sec:usr}), enforcing physical astrophysical coincidence across the detector network. The total model contains $\sim 2.9\times 10^6$ trainable parameters.

\subsubsection{Ranking statistic and post-hoc coincidence.}
\label{sec:usr}

To evaluate candidate GW events across the detector network without computationally prohibitive joint inference, we construct a closed-form ranking statistic from single-interferometer outputs. For each detector $d \in \{\mathrm{H}, \mathrm{L}\}$, the network outputs pre-softmax signal and noise logits, $z_{\mathrm{sig},d}$ and $z_{\mathrm{noise},d}$. Following Ref.~\cite{Schafer2022Training}, we evaluate the single-detector Unbounded Softmax Replacement (USR) statistic,

\begin{equation}
    s_d = z_{\mathrm{sig},d} - z_{\mathrm{noise},d}
        = \ln\!\left(\frac{p_{\mathrm{sig},d}}{1 - p_{\mathrm{sig},d}}\right),
\end{equation}

which represents the unconstrained signal-versus-noise log-odds.\ Unlike standard posterior probabilities bounded on $[0, 1]$ that saturate near unity for high-significance events, $s_d$ remains strictly monotonic and numerically stable in the deep low-false-alarm-rate (low-FAR) tail. 

Under the hypothesis of independent single-detector background triggers, we use an inclusive network detection rule, $p_{HL} = 1 - (1 - p_H)(1 - p_L)$, that maps these per-detector statistics into a coincident network log-odds:

\begin{equation}
  \mathrm{USR}_{HL} = \ln\!\left(e^{s_H} + e^{s_L} + e^{s_H + s_L}\right).
  \label{eq:usr}
\end{equation}

The cross-term $e^{s_H + s_L}$ naturally amplifies genuine coincident events, but the single-detector terms $e^{s_H}$ and $e^{s_L}$ permit loud, single-instrument non-Gaussian glitches to produce elevated ranking values.

To reject single-detector artifacts and enforce astrophysical consistency, we introduce a cross-detector frame-coherence statistic, $\mathcal{C}(f_H, f_L)$, derived from the localized temporal profiles produced by the frame head:

\begin{equation}
\mathcal{C}(f_H,f_L) = \max \Biggl[ 0,\, \max_{\delta\in\{-1,0,1\}}
\frac{
\displaystyle\sum_{j\in\mathcal{I}_{\delta}}
\widetilde{f}_{H,j}\widetilde{f}_{L,j-\delta}
}{
\displaystyle\max \left(
\sum_{j=1}^{64}\widetilde{f}_{H,j}^{\,2},\,
\sum_{j=1}^{64}\widetilde{f}_{L,j}^{\,2}
\right) + \epsilon
}
\Biggr],
\label{eq:framecoherence}
\end{equation}
where $\widetilde{f}_{d,j} = f_{d,j} - \overline{f}_d$ denotes the zero-mean frame profile across the $64$ tokens for detector $d \in \{\mathrm{H}, \mathrm{L}\}$, and $\epsilon = 10^{-6}$ prevents zero-energy division. The index set $\mathcal{I}_{\delta}$ restricts the sum strictly to overlapping tokens without periodic boundary wrap-around. 

Crucially, the token duration of $15.625\,\mathrm{ms}$ ($32$ samples at $2048\,\mathrm{Hz}$) cleanly resolves the light-travel propagation delay between LIGO Hanford and LIGO Livingston ($\le 10.0\,\mathrm{ms}$). Maximizing over temporal offsets $\delta \in \{-1, 0, 1\}$ allows for relative shifts of up to $\pm 15.625\,\mathrm{ms}$, fully spanning the physical light-travel time baseline. This penalizes single-detector noise glitches where only one interferometer exhibits an apparent signal profile.

The final detection ranking statistic therefore combines the network log-odds and the temporal coherence:
\begin{equation}
  \mathcal{R} = \mathrm{USR}_{HL} + w_\mathrm{c}\,\mathcal{C}(f_H,f_L),
  \label{eq:rank}
\end{equation}
adopting a fixed weight $w_\mathrm{c} = 4$, determined using ablation studies. Because $s_d$ and $f_d$ depend exclusively on single-interferometer strain, all model evaluations can be cached. 
A related strategy was adopted by McLeod et al.~\cite{McLeod2025BNS}, whose BNS search separates the network into detector-specific branches and a lightweight dense combiner. Time shifts are applied to the cached branch outputs so that only the inexpensive combiner must be re-evaluated for each slide. \castor\ follows the same general principle of decoupling the expensive detector-level inference from background generation. The implementations differ in that \castor\ combines the cached outputs using the closed-form statistic in Eq.~\eqref{eq:rank}, so each time slide requires only array recombination and evaluation of this expression, with no additional neural-network forward passes.

%Background estimation via time slides requires only evaluating Eq.~\eqref{eq:rank} on time-shifted cached arrays, bypassing repeated neural network forward passes and enabling computationally efficient large-scale background estimation. \\

\subsubsection{Training}
\label{sec:castor-train}
We train \castor\ using a hybrid curriculum that combines real-time, on-the-fly signal generation with pregenerated injections drawn from the MLGWSC-1 benchmark distributions (summarized in Table~I of Ref.~\cite{Schafer2023}). 

For the on-the-fly pipeline, all waveform simulation and signal-conditioning operations are performed directly on GPU using the \texttt{ml4gw} framework~\cite{ml4gw}. Each training instance is constructed from a 19 sec segment of real LIGO O3 noise: the first 16 secs are dedicated to estimating the background power spectral density (PSD), while the final 3 secs serve as the target window into which a simulated BBH may be injected. We estimate the PSDs independently for each interferometer via median averaging over 2 secs Hann-windowed segments with $50\%$ overlap, yielding a frequency resolution of $0.5$ Hz that effectively suppresses short, nonstationary noise transients. For this component of training, simulated signals are generated as 8 secs waveforms using the aligned-spin \texttt{IMRPhenomD} approximant~\cite{Husa2016,Khan2016}, drawing component masses uniformly over $[10, 50]\,\Msun$ ($m_1 \ge m_2$), dimensionless spin magnitudes uniformly in $[-0.999, 0.999]$, and network SNRs drawn from a power law over $[7, 20]$. 

To expose the network to richer waveform models, we incorporate \texttt{IMRPhenomXPHM} \cite{IMRPhenomXPHM}  injections in the pre-generated dataset. These waveforms accomodate precession and higher-order multipole modes and are evaluated from $f_{\min} = f_{\mathrm{ref}} = 20\,\mathrm{Hz}$ up to the Nyquist frequency of $1024\,\mathrm{Hz}$. The waveforms are projected onto H1 and L1 with isotropically sampled sky locations and orbital orientations, and rescaled to target network SNRs against the estimated PSDs. Identical time crops are applied to both detector channels before injecting the waveforms into O3b noise. The resulting 3 sec segments are conditioned with an instrument-specific, 2 sec finite-duration whitening filter with a $20$ Hz high-pass cutoff. We discard 1 sec from each boundary to remove filter corruption, and use the central 1 sec ($2048$ samples) as input to \castor. Simultaneously, the identical whitening operation is applied to a clean, noise-free copy of each injection, and the peak power of this conditioned template defines the target merger token for the frame-localization head.

We train the model under a joint objective: a regularized binary cross-entropy loss for the global signal-versus-noise classification head, combined with a Gaussian-profile binary cross-entropy loss (weighted by $w_{\mathrm{frame}} = 0.3$) for the per-token frame head. The optimization employs AdamW with a base learning rate of $5\times 10^{-4}$, weight decay of $10^{-2}$, a OneCycleLR schedule, and a batch size of $128$. Over the course of $500$ epochs, a curriculum smoothly increases the fraction of pregenerated \texttt{IMRPhenomXPHM} batches from $10\%$ to $30\%$, enabling the network to learn robust generic transient features before fine-tuning on precessing morphologies. 

\begin{figure}[t]
  \centering
  \includegraphics[width=\columnwidth]{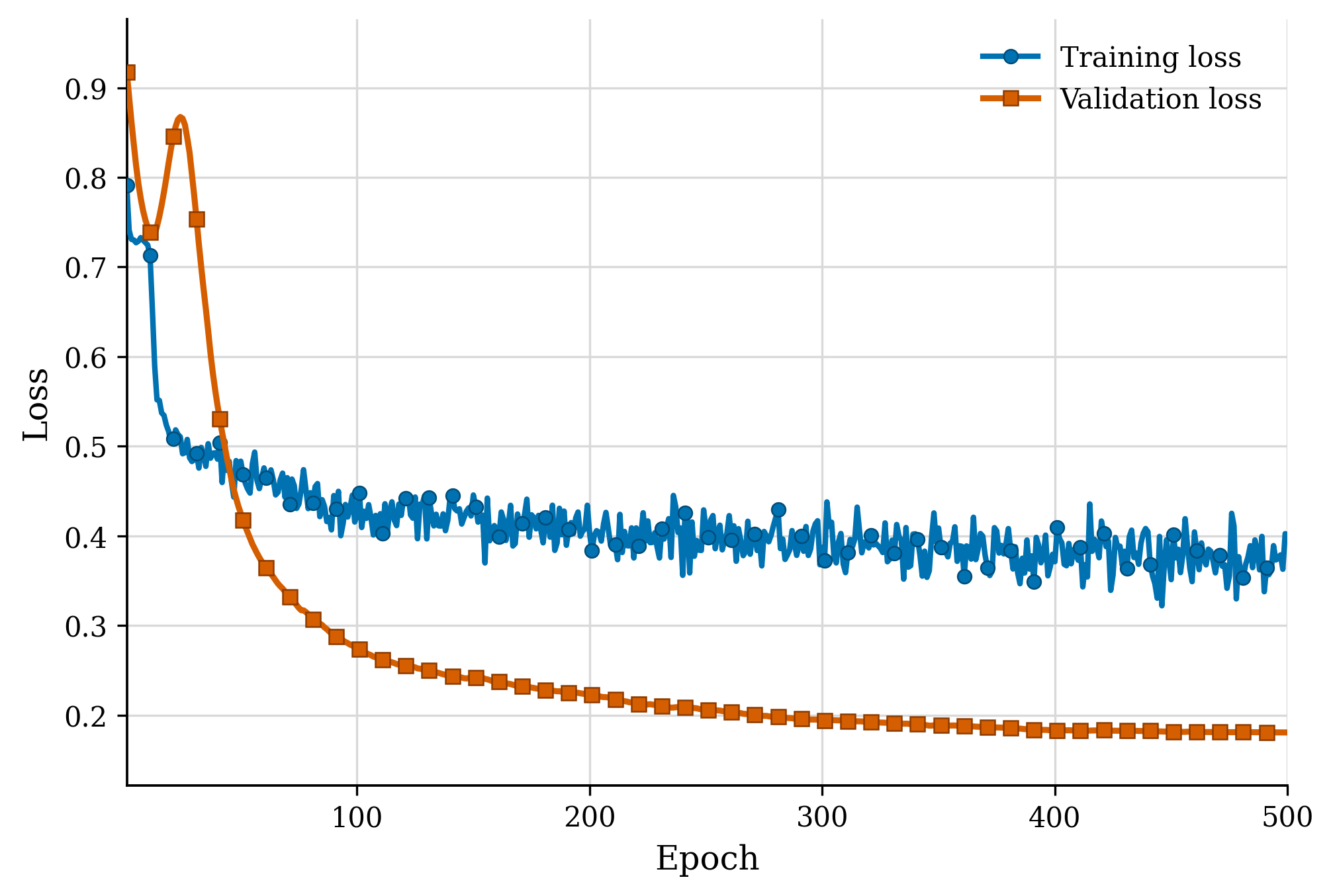}
  \caption{Training (blue) and validation (orange) loss for \castor\
  ($500$ epochs).  The validation loss (computed on a fixed pregenerated set
  with the exponential-moving-average weights) lies below the noisier
  on-the-fly training loss, and the network shows no sign of overfitting
  (Sec.~\ref{sec:castor-train}).}
  \label{fig:loss}
\end{figure}

The full training run completes in approximately $1.5\,\mathrm{hr}$ on a single $80\,\mathrm{GB}$ NVIDIA A100 GPU. Figure~\ref{fig:loss} presents the corresponding loss trajectories. The validation loss, computed on a fixed, pregenerated evaluation partition, lies consistently below the training loss and stabilizes without indication of overfitting. This offset arises because the training loss reflects stochastic, on-the-fly injections across an effectively infinite parameter space with fluctuating noise baselines, whereas the validation set provides a stationary benchmark. While training noise is sourced entirely from real O3 data, downstream generalizability is evaluated across both non-Gaussian O3a background (MLGWSC-1 dataset~4) and stationary colored Gaussian noise (dataset~3).

\subsection{\gww: an audio foundation model baseline}
\label{sec:gww}

To evaluate \castor\ against an existing transformer-based approach, we compare its performance to \gww~\cite{Chatterjee2024GWWhisper}, which adapts OpenAI's Whisper-tiny audio foundation model~\cite{Radford2023} ($\sim 39\times 10^6$ parameters) to GW detection. Rather than operating directly in the time domain, \gww\ converts each interferometer's whitened strain into a constant-$Q$ transform spectrogram ($Q$-scan) on GPU via \texttt{ml4gw}~\cite{ml4gw}. 

A lightweight front-end convolutional adapter with feature-wise linear modulation (FiLM) projects the single-detector spectrograms into the input dimension expected by the Whisper backbone. The underlying 4-layer, 6-head Whisper encoder remains frozen, while domain adaptation is achieved using weight-decomposed low-rank adaptation (DoRA)~\cite{Liu2024DoRA} targeting the attention output projections—updating only $\sim 1.5\%$ of the encoder weights. The resulting representations from H1 and L1 are concatenated and classified by an MLP head. 

Like \castor, \gww\ is evaluated on $1\,\text{sec}$ segments and trained on the on-the-fly O3 injection stream with network SNRs in $[7, 20]$. Full architectural and optimization specifications are provided in Ref.~\cite{Chatterjee2024GWWhisper}. \gww\ serves as a complementary baseline, allowing a direct comparison between a purpose-built, compact ($\sim 2.9\times 10^6$ parameter) time-domain transformer and a larger, repurposed audio foundation model operating on time--frequency representations.

\section{Data and evaluation}
\label{sec:data-evaluation}
\label{sec:data}
\label{sec:timeslides}
% ======================================================================

\subsection{MLGWSC-1 benchmark}
\label{sec:mlgwsc-evaluation}

We evaluate both networks on datasets~3 and~4 of MLGWSC-1~\cite{Schafer2023}. For each dataset, the challenge provides two independent one-month analyses: a background file containing only noise and a foreground file containing the same noise distribution with simulated BBH signals added. Each file has a live time of $T=2\,592\,000$ secs and contains coincident H1--L1 segments of at least $2$ hrs sampled at $2048~\mathrm{Hz}$. Dataset~3 contains Gaussian noise whose PSD is selected independently for each segment from a collection of O3a-derived PSDs. Dataset~4 instead contains real, coincident O3a strain, including non-Gaussian transients. The dataset-4 segments satisfy the GWOSC \texttt{DATA} flag, exclude times marked by the \texttt{CBC CAT1}, \texttt{CBC CAT2}, \texttt{CBC HW INJ}, or \texttt{BURST HW INJ} flags, and exclude $10$ secs around GWTC-2 events. The Livingston data are shifted by an independently chosen offset between $0$ and 240 secs in each segment to remove genuine inter-detector correlations.

The signal population is common to datasets~3 and~4. Injections are generated with \texttt{IMRPhenomXPHM} from $20~\mathrm{Hz}$, including precession and all available $(\ell,m)$ modes through $(4,4)$. Component masses are drawn from $7\leq m_2\leq m_1\leq50\,\Msun$, spin magnitudes from $[0,0.99]$ with isotropic orientations, and extrinsic parameters from the distributions in Table~I of Ref.~\cite{Schafer2023}. Waveform durations can reach approximately $20$ secs, and successive coalescences are separated by a random interval between $24$ and $30$ secs.

Both \castor\ and \gww\ use a common trigger-generation procedure but retain their model-specific preprocessing and ranking statistics. Each continuous segment of datasets~3 and~4 is whitened independently using the PyCBC Welch implementation \cite{PyCBC_software} (with its default median averaging) and Hann-windowed inverse-spectrum truncation, divided into overlapping 1 sec H1--L1 windows with a stride of $0.1$ sec, and evaluated in chronological order. Windows whose ranking statistic exceeds a first-level threshold are retained as triggers. Within each continuous segment, consecutive first-level triggers separated by no more than $0.35$ sec are assigned to the same cluster. This criterion permits a cluster to extend through a sequence of nearby windows. Each cluster is represented by the time and ranking statistic of its loudest member. 

No time slides are used for the MLGWSC-1 results. Following the challenge protocol~\cite{Schafer2023}, the FAR corresponding to a threshold $\mathcal{R}_{*}$ is measured directly from the noise-only background file,
\begin{equation}
  \mathrm{FAR}(\mathcal{R}_{*})
  =\frac{N_{\mathrm{bg}}(\mathcal{R}\geq\mathcal{R}_{*})}
  {2\,592\,000~\mathrm{s}}.
  \label{eq:mlgwsc-far}
\end{equation}
An injection is counted as found if the foreground output contains at least one event with the required ranking statistic whose reported time lies within the model-specific interval $\pm\Delta t$ of the injection time. The resulting found injections are converted to sensitive volume and sensitive distance using Eqs.~\eqref{eq:volume}--\eqref{eq:dsens} and the public MLGWSC-1 evaluation
code~\cite{Schafer2023}. 

%Time slides are performed only for \castor. Its ranking statistic is a closed-form combination of cached single-detector outputs, whereas \gww\ combines the two detector representations inside the network and must therefore be re-evaluated for every relative shift. This difference makes a \gww\ time-slide analysis prohibitively expensive for the present study: on one 80~GB NVIDIA A100 GPU, a one-month zero-lag foreground evaluation required approximately $3163~\mathrm{s}$ for \castor, whereas the one-month \gww\ background evaluation required approximately $18~\mathrm{h}$ (\(\sim20\times\) longer; Table~\ref{tab:timing} and Ref.~\cite{Chatterjee2024GWWhisper}). We therefore report the common zero-lag FAR comparison at $1~\mathrm{month}^{-1}$ for both networks and the extended $1~\mathrm{yr}^{-1}$ time-slide result only for \castor.

The architectural differences between the two networks translate directly into their computational costs. \castor\ evaluates each detector independently and combines the outputs after the network, allowing its single-detector representations to be cached. In contrast, \gww\ combines the two detector representations inside the network and must therefore evaluate the joint data streams through the full model. As detailed in Table~\ref{tab:timing}, training \castor\ takes approximately 1.5 hours, and performing inference on one month of MLGWSC-1 data requires approximately 0.9 hours on a single 80~GB NVIDIA A100 GPU. The same one-month inference for \gww\ takes approximately 18 hours. Because this base computational cost makes \gww\ roughly 20 times slower, re-evaluating the network for every relative time shift is prohibitively expensive. 

\begin{table}[t]
\caption{Model parameters and wall-clock computational costs for \castor\ and \gww\ using a single 80~GB NVIDIA A100 GPU. The inference time corresponds to the evaluation of one month of continuous MLGWSC-1 data. The \gww\ inference cost is taken from Ref.~\cite{Chatterjee2024GWWhisper}.}
\label{tab:timing}
\begin{ruledtabular}
\begin{tabular}{lcc}
Quantity & \castor & \gww \\
\colrule
Parameters                       & $\approx 2.9$\,M & $\approx 39$\,M \\
Training (per run)               & $\approx 1.5$\,hrs           & $\approx 8$\,hrs \\
1-month data inference           & $\approx 0.9$\,hrs     & $\approx 18$\,hrs  \\
\end{tabular}
\end{ruledtabular}
\end{table}

\subsection{O3b search and background estimation}
\label{sec:o3b-evaluation}

We additionally apply \castor\ to strain from the second half of the third observing run (O3b) and cross-match its zero-lag triggers against the confident BBH events in GWTC-3~\cite{GWTC3}. Following the data selection of Ref.~\cite{Zelenka2024}, we retain coincident, science-quality H1--L1 segments of at least $60$ secs, known hardware injections are not removed. This selection yields $2377$ segments with a total coincident
live time, $T_0=8\,228\,706~\mathrm{secs}\simeq95.3~\mathrm{days}$. A catalog event is considered recovered when the loudest clustered zero-lag trigger within $0.2~\mathrm{s}$ of its catalog time is selected.

\paragraph{Zero-lag background.}
``Zero-lag'' refers to the analysis of the simultaneous, unshifted detector data streams exactly as they were recorded, preserving true astrophysical coincidences rather than applying artificial relative time shifts. After matching triggers to catalog events, all remaining clustered zero-lag triggers are treated as false alarms. A candidate with ranking statistic $\mathcal{R}_*$, is assigned a FAR value given by Eq.~\ref{eq:far}. For the $T_0$ duration O3b data, the one-count FAR resolution is $1/T_0=0.315~\mathrm{month}^{-1}=3.84~\mathrm{yr}^{-1}$. Consequently this background is sufficient to evaluate the $1~\mathrm{month}^{-1}$ level. If no louder background event is present, the measured FAR is zero and we quote (and use for plotting) the one-count upper limit $1/T_0$.
%\begin{equation}
%  \mathrm{FAR}_0(\mathcal{R}_*)
%  =\frac{N_0(\mathcal{R}>\mathcal{R}_*)}{T_0},
%  \label{eq:o3b-zerolag-far}
%\end{equation}
%where $N_0$ is the number of unmatched zero-lag triggers louder than the
%candidate. 

\paragraph{\castor\ time-slide background.}
To extend the \castor\ measurement to $\mathrm{FAR}=1~\mathrm{yr}^{-1}$, we
construct an additional background using time slides. During the zero-lag pass, the single-detector
log odds and frame profiles $(s_H,s_L,f_H,f_L)$ are cached for every window.
The cached L1 sequence is then shifted relative to H1 by
\begin{equation}
  \tau_k=5k~\mathrm{s},\qquad k=1,\ldots,10,
  \label{eq:timeslides}
\end{equation}
giving ten statistically shifted realizations and approximately $10T_0\simeq2.61~\mathrm{yr}$ of background before the small edge correction described below. Each shift is applied independently within a continuous segment, without wrapping data across segment boundaries, and non-overlapping edge windows are discarded. The shifted H1--L1 pairs are ranked by recomputing Eq.~\eqref{eq:rank} from the cached quantities. The same thresholding and clustering algorithm is applied to zero-lag and time-slide triggers. The accumulated background live time is therefore given by,
\begin{equation}
  T_{\mathrm{bg}}
  =\sum_{k=1}^{10}\sum_j
    \max\!\left(0,L_j-\tau_k\right),
  \label{eq:timeslide-livetime}
\end{equation}
where $L_j=(N_j-1)\times0.1$ secs is the span between the first and last of the $N_j$ valid window times in the $j$th continuous segment. This definition accounts for the finite $1$ sec input window and for the non-overlapping windows discarded after a shift. For ten slides, the corresponding one-count resolution is approximately $1/T_{\mathrm{bg}}\simeq0.39~\mathrm{yr}^{-1}$, allowing FARs at the $1~\mathrm{yr}^{-1}$ level to be measured empirically. 

\section{Results}
\label{sec:results}
% ======================================================================

\begin{figure*}[t]
\centering
\includegraphics[width=\textwidth]{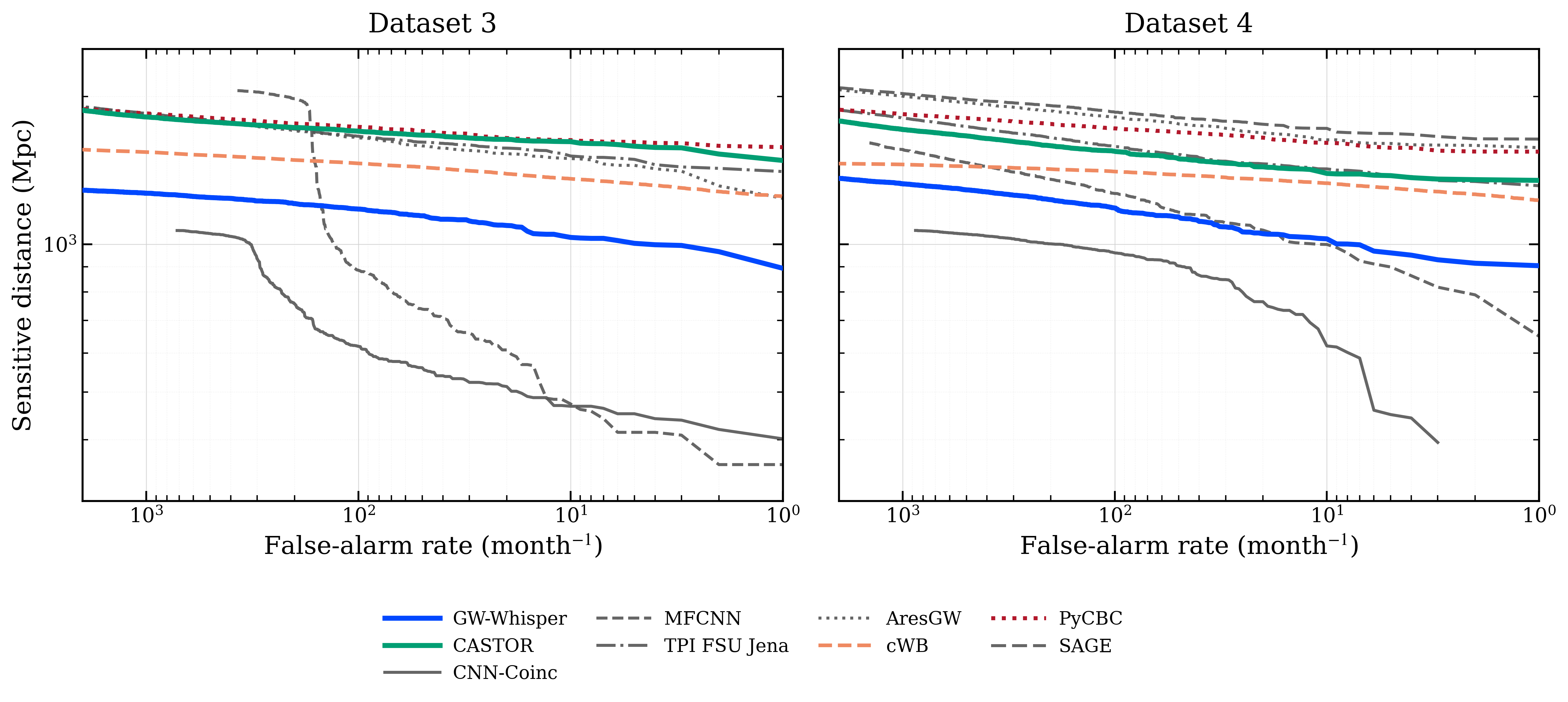}
\caption{Sensitive distance versus false-alarm rate for MLGWSC-1
dataset~3 (left) and dataset~4 (right), comparing \castor\ (green) and
\gww\ (blue) with the other MLGWSC-1 learned searches and the standard
PyCBC and cWB analyses. Where available, the most recent published results
from each pipeline on these datasets are shown. \castor\ is among the most
sensitive pipelines on dataset~3 and remains substantially more sensitive
than \gww\ on both datasets.}
\label{fig:sens}
\end{figure*}

Figure~\ref{fig:sens} shows the sensitive distance [Eqs.~\eqref{eq:volume}--\eqref{eq:dsens}] as a function of FAR for datasets~3 and~4. The comparison includes the learned searches CNN-Coinc, MFCNN~\cite{Wang2020}, TPI-FSU Jena~\cite{Zelenka2024}, AResGW~\cite{Nousi2023}, and SAGE~\cite{sage}, together with the standard PyCBC~\cite{Usman2016} and cWB~\cite{Klimenko2016} analyses. For pipelines with updated evaluations on these datasets, the latest published results are used.

At a FAR of $1\ \mathrm{month}^{-1}$, \castor\ reaches sensitive distances of $1481.96\ \mathrm{Mpc}$ on dataset~3 and $1349.72\ \mathrm{Mpc}$ on dataset~4, compared with $893.70\ \mathrm{Mpc}$ and $904.44\ \mathrm{Mpc}$, respectively, for \gww. Thus, \castor\ improves the sensitive distance by factors of approximately $1.66$ and $1.49$, corresponding to factors of approximately $4.6$ and $3.3$ in sensitive volume.

We find that SAGE and AResGW exceed \castor\ at $1\ \mathrm{month}^{-1}$ on dataset~4. Both process the two detector streams jointly and can learn cross-detector consistency within the network, making them coherent searches in the operational sense used here. Their improved sensitivity is therefore consistent with the additional information available to a joint detector model. In contrast, \castor\ encodes each detector separately and combines the cached outputs through its coincidence statistic. This design permits time-slide backgrounds to be generated without additional network evaluations, whereas joint models generally require the shifted detector pairs to be evaluated again for every slide. The modest sensitivity trade-off therefore provides a substantial reduction in the computational cost of estimating low FARs.

\begin{figure*}[t]
\centering
\includegraphics[width=0.48\textwidth]{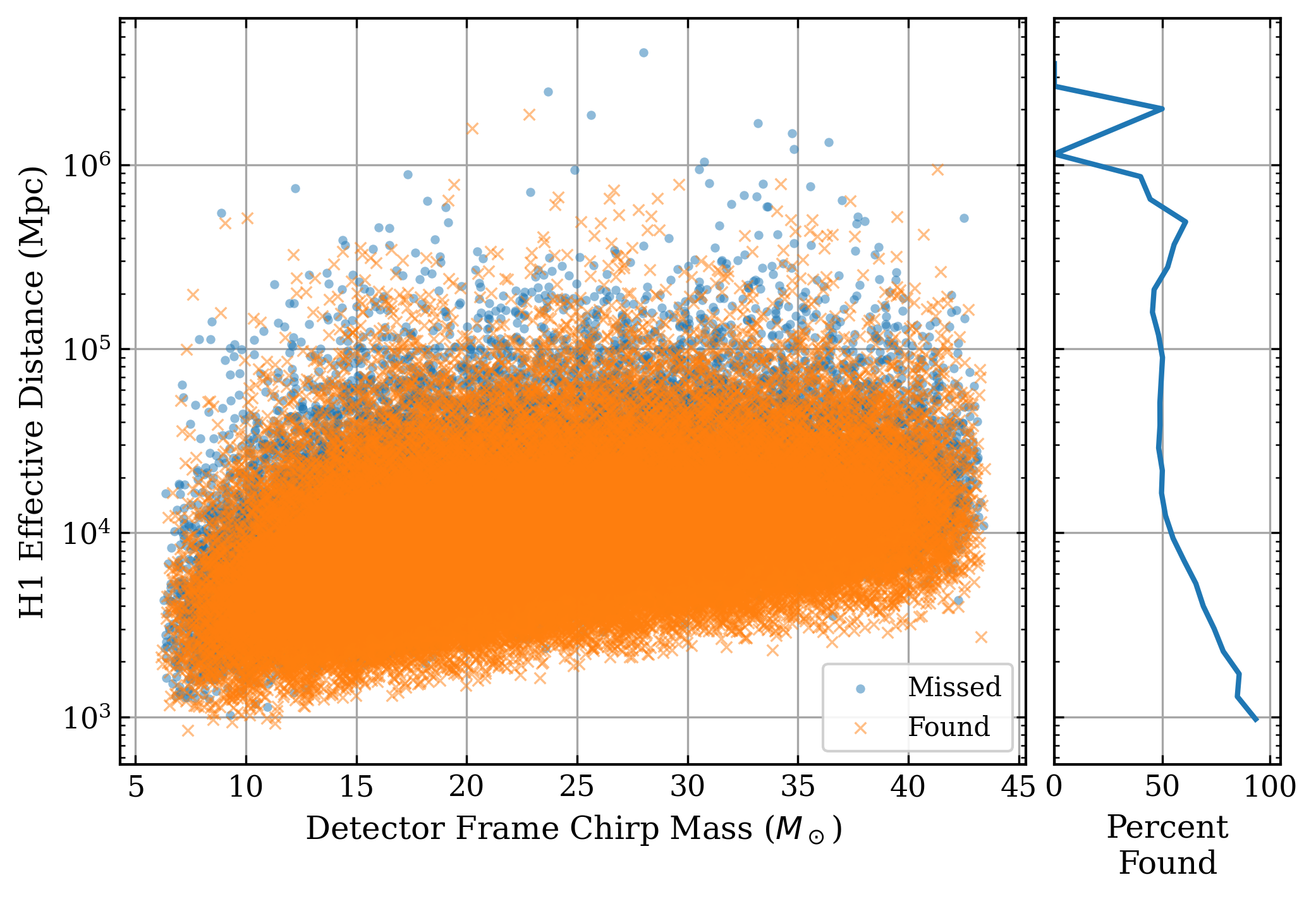}
\hfill
\includegraphics[width=0.48\textwidth]{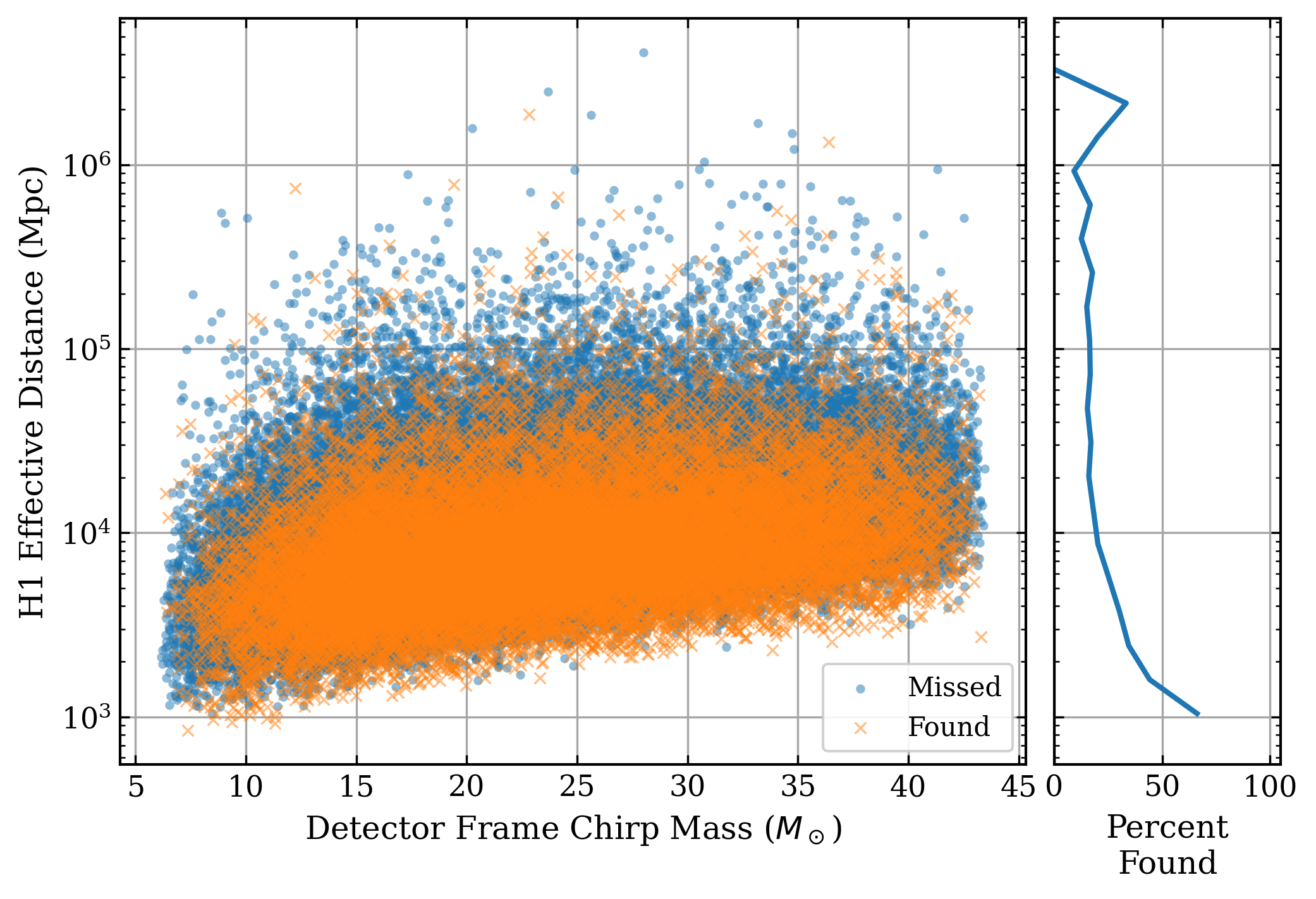}
\caption{Found (orange) and missed (blue) dataset-4 injections in the detector-frame chirp-mass and H1 effective distance plane, together with the recovered fraction versus H1 effective distance, for CASTOR (left) and GW-Whisper (right). The smallest-distance bins are dominated by low-mass systems. CASTOR's recovered fraction increases for these nearby signals, whereas that of GW-Whisper decreases sharply.}
\label{fig:foundmissed}
\end{figure*}

Figure~\ref{fig:foundmissed} uses the dataset~4 results to provide further insight into the sensitivity difference between \castor\ and \gww. In each logarithmically spaced distance bin, the recovered fraction is
\begin{equation}
\epsilon_i =
\frac{N_{\mathrm{found},i}}
{N_{\mathrm{found},i}+N_{\mathrm{missed},i}}.
\end{equation}

Unlike the luminosity distance, the effective distance accounts for the projection of the signal onto the detector and is therefore more closely related to the amplitude present in the strain data. Both models preferentially recover injections at smaller effective distances, but \castor\, maintains a markedly higher recovery fraction over most of the sampled range. GW-Whisper exhibits a substantially sharper loss of efficiency with increasing effective distance, demonstrating its lower sensitivity to weakly observed signals.

A primary driver of the disparity between the performances of \castor\ and \gww\ lies in the input representation and its associated normalization. \gww\ converts whitened strain into time--frequency spectrograms via a constant-$Q$ transform and standardizes each scan by its per-instance mean and standard deviation. While this stabilizes neural network training across varying noise floors, it discards the absolute amplitude information that distinguishes nearby, high-SNR chirps from weak background fluctuations. For exceptionally loud signals at small distances, the prominent chirp track heavily skews these global statistics, altering the apparent noise floor and creating out-of-distribution spectrogram patterns that \gww\ can misclassify as noise or unmodelled non-Gaussian transients. In contrast, \castor\ operates directly on whitened time-domain strain, preserving relative amplitude profiles and phase coherence across the entire dynamic range.

Furthermore, \gww\ faces representational limitations arising from domain transfer. Because its underlying Whisper backbone was pretrained on human speech, its attention mechanisms are intrinsically optimized for acoustic features such as formants, harmonic structure, and vocal pauses rather than the strictly deterministic, continuous frequency sweeps governed by general relativity. Adapting only $\sim 1.5\%$ of the encoder weights via low-rank adaptation proves insufficient to fully overcome these acoustic inductive biases or to capture faint, highly dispersed chirp tracks in non-stationary noise. This challenge is compounded for lower-mass systems ($\mathcal{M}_c \approx 6$--$9\,M_\odot$), which dominate the nearest distance bins. These signals remain in band across many cycles, sweeping through frequency space at rates that suffer under the fixed resolution trade-offs of static time--frequency binning. By learning adaptive, multi-scale temporal filter banks directly from raw strain through its strided convolutional front end, \castor\ avoids these transform-induced distortions and maintains robust recovery across both extended inspirals and short, massive coalescences.

\subsection{Background validation}
\label{sec:farval}

\begin{figure}[t]
\centering
\includegraphics[width=\columnwidth]{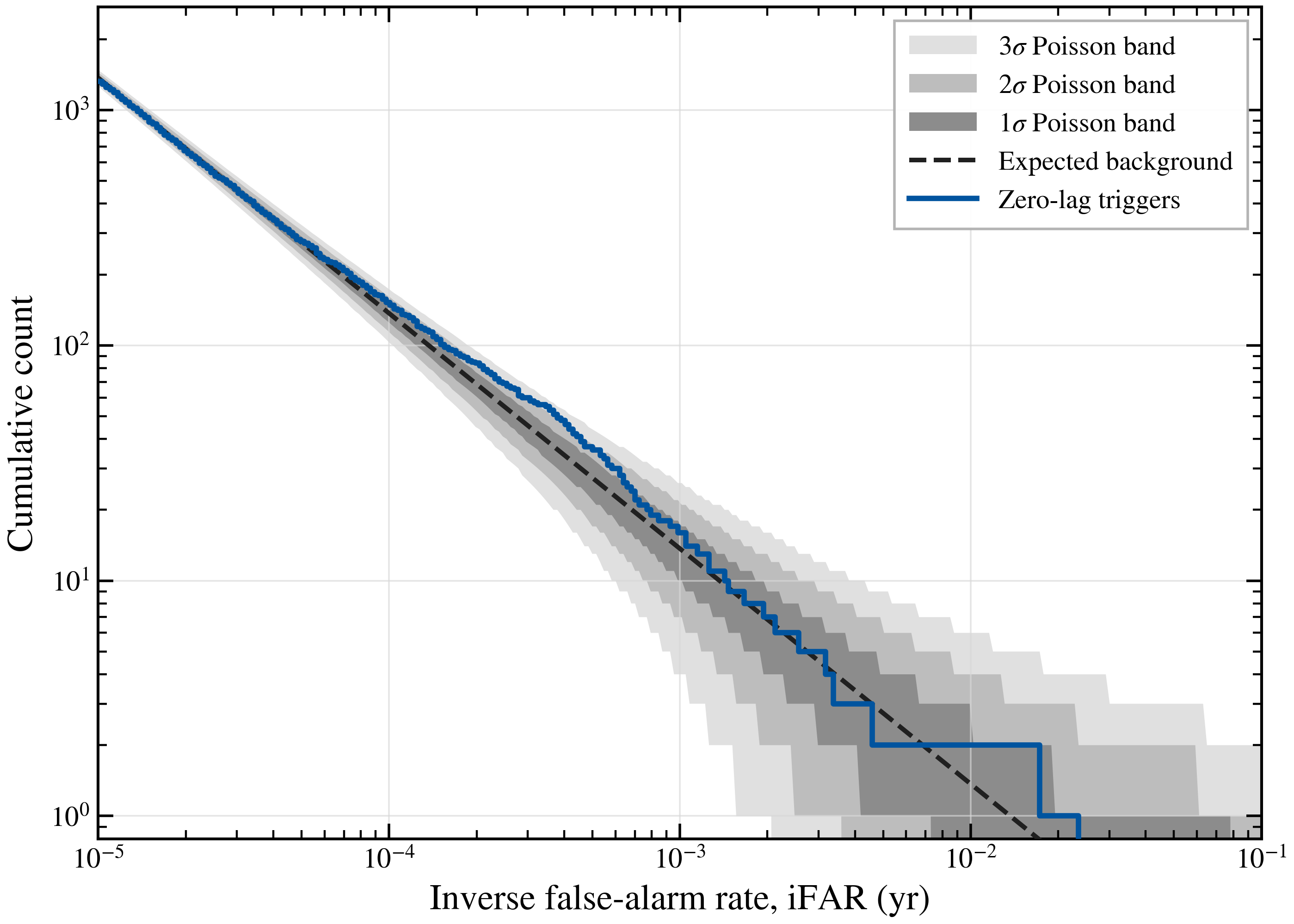}
\caption{Background validation for \castor\ using independent O3a data. The cumulative number of zero-lag triggers (blue) is shown as a function of inverse false-alarm rate and compared with the time-slide expectation (dashed), together with its $1$--$3\sigma$ Poisson intervals (grey). The agreement indicates that the time-slide background provides a statistically consistent estimate of the false-alarm distribution.}
\label{fig:farval}
\end{figure}

The computational advantage of the \castor\ time-slide procedure is useful only if the resulting background provides a statistically faithful description of real coincident noise. We test this in Fig.~\ref{fig:farval} using two disjoint stretches of O3a data. The zero-lag distribution is constructed from five days of coincident H1--L1 data selected between GPS times $1239392735$ and $1240303634$. The background is constructed from a separate five days of coincident data between GPS times $1238166018$ and $1238997902$. The nonoverlapping intervals ensure that the zero-lag triggers are statistically independent of the detector data used to estimate the background.

We perform 300 relative time slides of the background data, providing roughly four years of nominal background exposure before accounting for samples discarded at segment boundaries. For each inverse-FAR threshold, the observed cumulative zero-lag count is compared with the expected count obtained by multiplying the corresponding FAR by the zero-lag live time. The grey regions show the $1$--$3\sigma$ Poisson fluctuations about this expectation. The zero-lag distribution follows the time-slide prediction and remains consistent with these intervals across the evaluated IFAR range. This agreement demonstrates that the time-slide procedure accurately reproduces the coincident-noise tail and supports the assignment of empirical false-alarm rates to \castor\ candidate events.

\subsection{Application to O3b}
\label{sec:o3b}

\begin{figure}[t]
\centering
\includegraphics[width=\columnwidth]{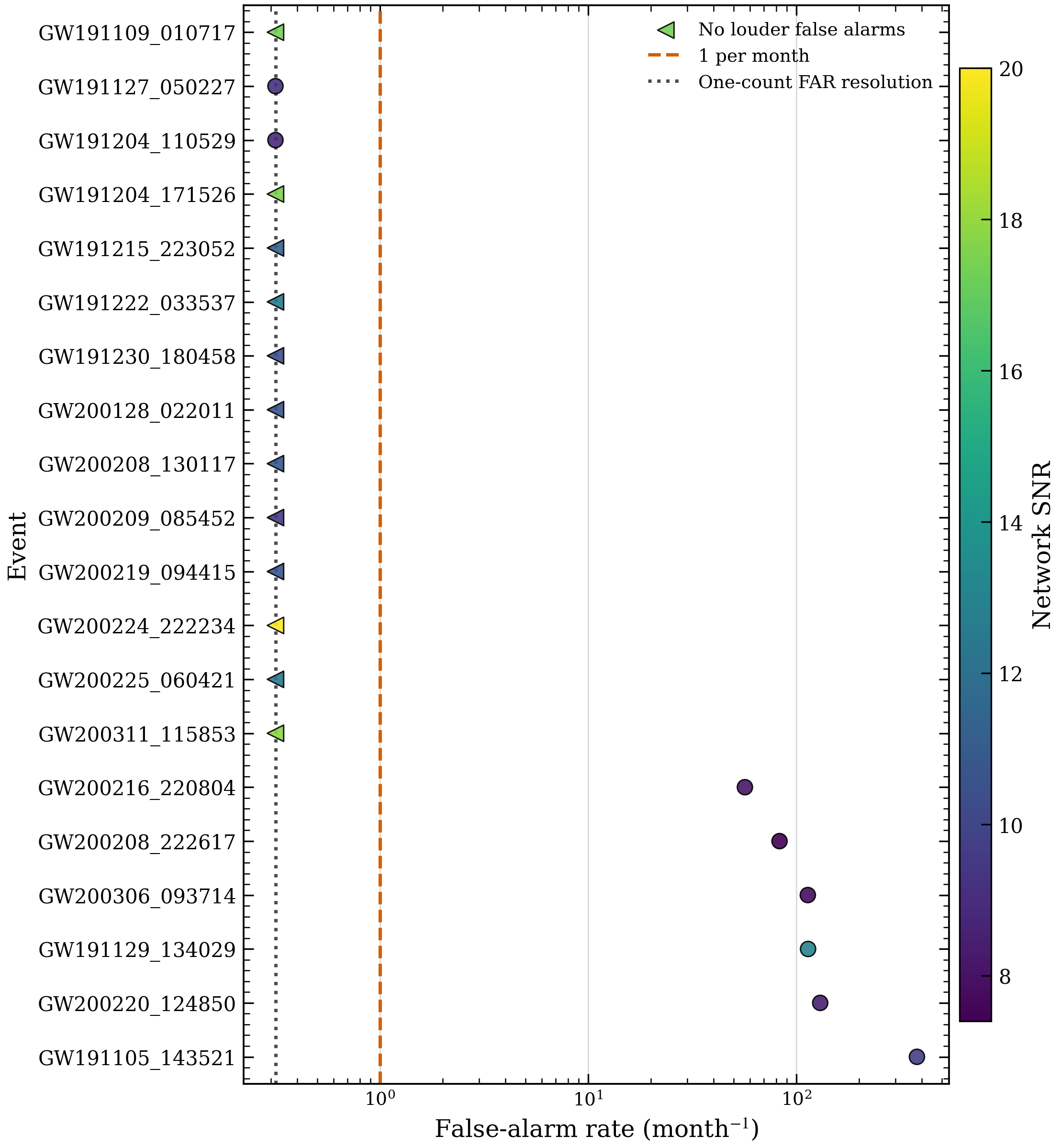}
\caption{False-alarm rates, expressed per month, assigned by \castor\ to the recovered confident O3b BBH events in GWTC-3. Events are colour-coded by their catalogue network SNR~\cite{GWTC3}. Left-pointing triangles denote events louder than every background event, for which the FAR is limited by the one-count resolution $1/T_\mathrm{bg}$ (dotted line). The dashed line marks $1\ \mathrm{month}^{-1}$. \castor\ recovers 14 events with $\mathrm{FAR}<1\ \mathrm{month}^{-1}$.}
\label{fig:o3bcastormonth}
\end{figure}

We apply \castor\ to O3b data and cross-match the resulting triggers against the 31 confident O3b BBH events in GWTC-3~\cite{GWTC3}, subject to the data availability requirements described in Sec.~\ref{sec:timeslides}. That section also describes the time-slide construction and the background exposure needed to resolve a FAR of $1\ \mathrm{month}^{-1}$. The complete zero-lag and time-slide analysis required around 5 hours on a DGX 80 GB A100 GPU.

%Figure~\ref{fig:o3bcastormonth} shows the empirical FAR assigned to each recovered event. \castor\ recovers 14 confident events with $\mathrm{FAR}<1\ \mathrm{month}^{-1}$. Several of the loudest events, including GW200224\_222234, GW200311\_115853, GW191109\_010717, and GW191204\_171526, are louder than every event in the background. Their reported FARs are therefore limited by the one-count resolution $1/T_\mathrm{bg}$. A further 6 events are recovered with FARs between $1$ and $10^{3}\ \mathrm{month}^{-1}$. We classify an event as missed when no zero-lag trigger falls within the $0.2$ sec coincidence window of the catalogue merger time, or when its loudest associated trigger is assigned a false-alarm rate exceeding $10^{3}\ \mathrm{month}^{-1}$. Such events are marked ``missed'' in Table~\ref{tab:o3b}. The per-year FARs obtained from time-slides are listed alongside the per-month values in Table~\ref{tab:o3b}. We found that the loudest events reached $\mathrm{FAR}<1\ \mathrm{yr}^{-1}$ from analyzing the empirical background.

Figure~\ref{fig:o3bcastormonth} shows the empirical FAR assigned to each recovered event. We define a confident recovery as having a $\mathrm{FAR}\le 1\ \mathrm{month}^{-1}$. \castor\ confidently recovers 14 events meeting this threshold. Several of the loudest events, including GW200224\_222234, GW200311\_115853, GW191109\_010717, and GW191204\_171526, are louder than every event in the background. Their reported FARs are therefore limited by the one-count resolution $1/T_\mathrm{bg}$. To maintain a direct comparison with the evaluation conventions of Ref.~\cite{Zelenka2024}, we track triggers up to a FAR of $10^{3}\ \mathrm{month}^{-1}$. A further 6 events are recovered with FARs between $1$ and $10^{3}\ \mathrm{month}^{-1}$. Because this rate is too high for astrophysical discovery, we categorize these strictly as low-significance events. We classify an event as completely missed when no zero-lag trigger falls within the $0.2$ sec coincidence window of the catalogue merger time, or when its loudest associated trigger is assigned a false-alarm rate exceeding $10^{3}\ \mathrm{month}^{-1}$. Such events are marked ``missed'' in Table~\ref{tab:o3b}. The per-year FARs obtained from time-slides are listed alongside the per-month values in Table~\ref{tab:o3b}. We found that the loudest events reached $\mathrm{FAR}<1\ \mathrm{yr}^{-1}$ from analyzing the empirical background.

Table~\ref{tab:o3b} summarizes the recovery of the confident catalogue events. Ref.~\cite{Zelenka2024} separates events into three groups according to whether the $90\%$ credible intervals of their component masses lie fully inside, overlap the boundaries of, or lie fully outside the $[10,50]\,\Msun$ training range. For the simpler binary classification used here, we instead use the posterior median component masses: an event is labelled ``in'' when both medians lie in $[10,50]\,\Msun$ and ``out'' otherwise. Because this label considers only component masses, it should not be interpreted as a complete characterization of whether an event is in distribution with respect to all training parameters. The events not confidently recovered by \castor\ fall into the following groups:

\begin{enumerate}
\item \textbf{Events outside the nominal component-mass range.}
Most missed events have at least one posterior median component mass outside $[10,50]\,\Msun$. These are predominantly neutron-star--black-hole or low-secondary-mass systems, including
GW191219\_163120, GW200115\_042309, GW200210\_092254, GW200202\_154313, and GW191129\_134029. GW200220\_061928 instead lies above the nominal mass range. These events were also missed by all six networks considered in Ref.~\cite{Zelenka2024}, indicating a common limitation associated with the training mass coverage.

\item \textbf{Low-SNR events.}
Some events have among the lowest matched-filter SNRs in the confident catalogue. GW200322\_091133, whose median component masses lie within the nominal range, has $\rho_\mathrm{MF}=6.0$. GW200308\_173609 has $\rho_\mathrm{MF}=7.1$, although its median primary mass lies above the nominal range. Both events are missed, consistent with reduced search sensitivity near the detection threshold, potentially compounded by limited mass coverage for the latter event.

\item \textbf{Events excluded because of data quality.}
Four events, GW200302\_015811, GW200129\_065458, GW200112\_155838, and GW191216\_213338, were excluded because of insufficient data quality in one or both detectors and are therefore absent from our analysis.
\end{enumerate}

The additional events GW200201\_203549, GW200214\_224526,
GW200219\_201407, and GW200311\_103121 appearing in the search outputs are
classified as marginal GWTC-3 candidates and are therefore not included in
the count of 31 confident events or in the confident-event recovery rate.

Overall, the O3b non-detections are dominated by sources outside the nominal component-mass range, events close to the detection threshold, and intervals excluded by the data-quality selection. We do not claim that \castor\ matches the confident-event recovery of matched-filter searches. The purpose of this comparison is to evaluate its performance relative to learned searches using the same O3b data selection and a transparent, mass-based classification.

\begin{table*}[t]
\caption{O3b GWTC-3 confident BBH events and their recovery by \castor. ``$\rho_\mathrm{MF}$'' is the catalogue matched-filter network SNR~\cite{GWTC3}, and ``FAR'' is the false-alarm rate assigned by the empirical time-slide background, given both per month and per year. \castor\ confidently recovers $14$ events at $\mathrm{FAR}\le1\ \mathrm{month}^{-1}$. Events recovered with a FAR between $1$ and $10^{3}\ \mathrm{month}^{-1}$ are categorized as low-significance. An entry of ``missed'' denotes an event for which no zero-lag trigger falls within the $0.2$ sec coincidence window, or whose loudest associated trigger has $\mathrm{FAR}>10^{3}\ \mathrm{month}^{-1}$. ``Mass range'' gives a descriptive classification based on the posterior median component masses: ``in'' means that both medians lie within $[10,50]\,\Msun$, while ``out'' means that at least one lies outside this interval. This mass-based label is not a complete characterization of distribution shift across all source and noise parameters. The O3b data selection follows Ref.~\cite{Zelenka2024}. Events are ordered by decreasing significance.}
\label{tab:o3b}
\begin{ruledtabular}
\begin{tabular}{lcccc}
Event & $\rho_\mathrm{MF}$ & FAR (month $^{-1}$) & FAR (year$^{-1}$) & Mass range \\
\colrule
GW200224\_222234 & 20.0 &  $< 0.31$ & $< 0.39$ & in \\
GW200311\_115853 & 17.8 &  $< 0.31$ & $< 0.39$ & in \\
GW191109\_010717 & 17.3 & $< 0.31$ & $< 0.39$ & out \\
GW191204\_171526 & 17.5 & $< 0.31$ & $< 0.39$ & out \\
GW200225\_060421 & 12.5 & $< 0.31$ & $< 0.39$ & in \\
GW191222\_033537 & 12.5 & $< 0.31$ & $< 0.39$ & in \\
GW191215\_223052 & 11.2 & $< 0.31$ & $< 0.39$ & in \\
GW200128\_022011 & 10.6 & $< 0.31$ & $< 0.39$ & in \\
GW200209\_085452 & 9.6  & $< 0.31$ & $< 0.39$ & in \\
GW200208\_130117 & 10.8 & $< 0.31$ & $0.39$ & in \\
GW200219\_094415 & 10.7 & $< 0.31$ & $0.39$ & in \\
GW191230\_180458 & 10.4 & $< 0.31$ & $0.39$ & in \\
%\colrule
%\multicolumn{4}{c}{recovered at higher FAR / missed, see
%Sec.~\ref{sec:o3b}}\\
GW191204\_110529 & 8.8  & $0.31$ & $1.15$ & in \\
GW191127\_050227 & 9.2  & $0.31$ & $2.68$ & out \\
GW200216\_220804 & 8.1  & $56.69$ & $753.59$ & out \\
GW200208\_222617 & 7.4  & $83.16$ & $1068.83$ & out \\
GW200306\_093714 & 7.8  & $113.71$ & $1410.92$ & in \\
GW191129\_134029 & 13.1 & $114.03$ & $1425.11$ & out \\
GW200220\_124850 & 8.5  & $130.40$ & $1662.12$ & in \\
GW191105\_143521 & 9.7  & $380.20$ & $4617.41$ & out \\
GW191103\_012549 & 8.9  & missed & missed & out \\
GW200202\_154313 & 10.8 & missed & missed & out \\
GW200316\_215756 & 10.3 & missed & missed & out \\
GW200308\_173609 & 7.1  & missed & missed & out \\
GW200322\_091133 & 6.0  & missed & missed & in \\
GW191126\_115259 & 8.3  & missed & missed & out \\
GW200115\_042309 & 11.6 & missed & missed & out \\
GW200220\_061928 & 7.2  & missed & missed & out \\
GW191113\_071753 & 7.9  & missed & missed & out \\
GW191219\_163120 & 9.1  & missed & missed & out \\
GW200210\_092254 & 8.4  & missed & missed & out \\
\end{tabular}
\end{ruledtabular}
\end{table*}

% ======================================================================
\section{Discussion}
\label{sec:discussion}
% ======================================================================

We have presented \castor, a fast and accurate transformer-based search for BBHs: a compact per-detector time-domain transformer trained from scratch whose coincident, post-hoc ranking statistic makes time-slide background estimation
more than an order of magnitude cheaper than for a jointly-evaluated network, because each slide reuses the cached single-detector outputs and requires no additional neural-network calls. The resulting FAR is statistically well
calibrated and needs no extrapolation, at a modest, quantified cost in sensitivity relative to coherent analyses. On MLGWSC-1 datasets~3 and~4, \castor\ is among the most sensitive learned pipelines, and on O3b it recovers
the majority of in-range confident GWTC-3 events, with its non-detections dominated by events outside its training domain or near threshold. We benchmarked \castor\ against  \gww, a domain-adapted audio foundation model, and found that \castor\ outperforms it by roughly a factor of two in sensitive distance and is roughly 20 times faster at inference. We make no claim of state-of-the-art sensitivity or production readiness of \castor\ at this stage. Rather, these results demonstrate a lightweight, carefully designed time-domain transformer that offers a practical route to cheap, trustworthy background estimation for coincident
deep-learning searches.

Our results carry three central messages. The first is methodological: \castor\ demonstrates a fast, accurate transformer-based BBH search whose coincident, post-hoc combination design overcomes a major computational bottleneck in background estimation. As established by the inference timings in Sec.~\ref{sec:mlgwsc-evaluation}, generating the multi-year background exposure required to confidently evaluate the loudest O3b events using a joint-detector network would demand hundreds of GPU-hours. By decoupling the neural-network evaluation from the time-slide recombination, \castor\ allows the total background live time to be scaled up at a negligible arithmetic cost. This permits the assignment of fully empirical FARs below $1\,\mathrm{yr}^{-1}$ without resorting to the exponential extrapolation often required by coherent machine-learning pipelines~\cite{Chatterjee2024GWWhisper}. This speed-up quantifies a fundamental design trade-off: a coincident statistic intrinsically discards the relative phase information exploited by coherent analyses, sacrificing a degree of sensitivity (Fig.~\ref{fig:sens}) in exchange for background estimation that is computationally cheaper. For operational scenarios where vast backgrounds must be generated repeatedly such as iterative threshold tuning, ranking-statistic development, or low-latency alert generation, this provides a highly practical and transferable recipe.

A second message, and a finding of the transformer-to-transformer comparison, is that on identical benchmarks the compact, purpose-built time-domain transformer trained from scratch substantially outperforms the much larger repurposed audio foundation model (Fig.~\ref{fig:sens}), and does so with roughly an order of magnitude fewer parameters. This outcome is expected rather than surprising: an audio foundation model is not specialised to GW morphology, so it should not be read as an argument against transfer learning or foundation models for GW science. \gww\ itself demonstrates rapid convergence, strong glitch representations, and parameter-efficient adaptation~\cite{Chatterjee2024GWWhisper}, and a transformer pre-trained on GW data and fine-tuned for a GW task could plausibly perform well. Our result rather shows that, for BBH detection at fixed FAR, the inductive bias of a time-domain representation and an architecture co-designed with the coincidence structure of the problem currently outweighs the transfer of features from audio.

Third, the high-SNR behaviour is a cautionary tale about input representations: a spectrogram front end inherited from a fixed-dynamic-range audio pipeline can fail precisely on the loudest, most significant events, whereas a time-domain network retains amplitude information and degrades gracefully. Any learned search intended for real use should be explicitly stress-tested at SNRs well above its training range.

The principal limitations of this study are the following: as noted above, \castor's sensitivity, while competitive with the best learned pipelines on MLGWSC-1, remains below the best coherent and matched-filter analyses due to the deliberate trade of sensitivity for background-estimation speed. The training mass and SNR ranges ($[10,50]\,\Msun$, $\rho\in[7,20]$) restrict the domain of validity as events outside these ranges (Sec.~\ref{sec:o3b}) are not expected to be recovered. The on-the-fly training waveforms use the dominant-mode, aligned-spin \texttt{IMRPhenomD} approximant, as \texttt{ml4gw} does not currently implement GPU-accelerated waveform generation that supports higher-order modes. While the test signals include precession and higher-order modes, the pregenerated-waveform curriculum partially mitigates this mismatch. Finally, we restrict attention to BBH systems and two detectors. Extensions to lower masses (longer signals, requiring longer input windows), to three or more detectors, and to a learned, but still separable, combination of the single-detector statistics are natural directions for future work.

% ======================================================================
\section{Conclusion}
\label{sec:conclusion}
% ======================================================================
We have presented \castor, a fast and accurate transformer-based search for BBHs: a compact per-detector time-domain transformer trained from scratch whose coincident, post-hoc ranking statistic makes time-slide background estimation
more than an order of magnitude cheaper than for a jointly-evaluated network, because each slide reuses the cached single-detector outputs and requires no additional neural-network calls. The resulting FAR is statistically well
calibrated and needs no extrapolation, at a modest, quantified cost in sensitivity relative to coherent analyses. On MLGWSC-1 datasets~3 and~4, \castor\ is among the most sensitive learned pipelines, and on O3b it recovers
the majority of in-range confident GWTC-3 events, with its non-detections dominated by events outside its training domain or near threshold. We benchmarked \castor\ against  \gww, a domain-adapted audio foundation model, and found that \castor\ outperforms it by roughly a factor of two in sensitive distance and is roughly 20 times faster at inference. We make no claim of state-of-the-art sensitivity or production readiness of \castor\ at this stage. Rather, these results demonstrate a lightweight, carefully designed time-domain transformer that offers a practical route to cheap, trustworthy background estimation for coincident
deep-learning searches.

\section*{Code Availability}
The \castor\ pipeline is implemented in \textsc{Python} (version $\ge 3.10$) using \textsc{PyTorch} (version $\ge 2.0$) and \texttt{ml4gw}. The package is publicly available on the Python Package Index (PyPI) at \url{https://pypi.org/project/castor-gw/}. Complete source code, training configurations, and reproduction scripts are hosted on GitHub at \url{https://github.com/chayanchatterjee/castor}.

\begin{acknowledgments}
K.J. acknowledges support from NSF CAREER \#2544531. C.C. and K. J.’s work was supported in part by the Lunar Labs Initiative at Vanderbilt University, which is funded by Cornelius Vanderbilt Dean's Faculty Fellowship and Scaling Grant from Vanderbilt Office of the Vice Provost for Research and Innovation. This research was undertaken with the support of compute grant and resources located at Vanderbilt University, USA. This material is based upon work supported by NSF's LIGO Laboratory which is a major facility fully funded by the National Science Foundation. This research used data obtained from the Gravitational Wave Open Science Center (https://www.gw-openscience.org), a service of LIGO Laboratory, the LIGO Scientific Collaboration and the Virgo Collaboration. LIGO is funded by the U.S. National Science Foundation. Virgo is funded by the French Centre National de Recherche Scientifique (CNRS), the Italian Istituto Nazionale della Fisica Nucleare (INFN) and the Dutch Nikhef, with contributions by Polish and Hungarian institutes.
\end{acknowledgments}

% ======================================================================
%  FIGURES
% ======================================================================

%\begin{figure}[t]
%  \centering
%  \includegraphics[width=\columnwidth]{O3b_results/o3b_event_fars_GW-Whisper_per_month.png}
%  \caption{As Fig.~\ref{fig:o3bcastormonth}, but for \gww, using the same empirical
%  time-slide background construction.  With an identical, non-extrapolated background, \gww\
%  confidently recovers substantially fewer O3b events at
%  $\mathrm{FAR}<1\,\mathrm{month}^{-1}$ than \castor.}
%  \label{fig:o3bgww}
%\end{figure}

% ======================================================================
%  TABLES
% ======================================================================

% ======================================================================
\bibliographystyle{apsrev4-2}
\bibliography{apssamp}
% ======================================================================

\end{document}